\documentclass[letter,useAMS,usenatbib]{mn2e}

\usepackage{amsmath}
\usepackage{amssymb,amsfonts}
\usepackage{graphicx,epsfig,graphics}
\usepackage{graphics}
\usepackage{ulem}
\usepackage{caption}
\usepackage[ruled,vlined]{algorithm2e}
\providecommand{\SetAlgoLined}{\SetLine}

\usepackage{longtable}
\usepackage{subfigure}
\newcommand{\bt}{\mbox{\boldmath{$\theta$}}}

\newcommand{\bg}{\mbox{\boldmath{$\gamma$}}}

\newcommand{\vir}{\rm vir}
\DeclareMathOperator\erf{erf}

\DeclareMathOperator\MAD{MAD}

\SetKwInput{KwInitialization}{Initialization}
\SetKwInput{KwParameters}{Parameters}
\begin{document}

\label{firstpage}
\title[GLIMPSE: Lensing Reconstructions Using Sparsity]{GLIMPSE: Accurate 3D weak lensing reconstructions using sparsity} 
\author[A. Leonard, F. Lanusse, J.-L. Starck] {Adrienne
  Leonard\thanks{Email: adrienne.leonard@cea.fr}, Fran\c{c}ois Lanusse\thanks{Email: francois.lanusse@cea.fr}, Jean-Luc Starck\\Laboratoire AIM, UMR CEA-CNRS-Paris 7, Irfu, Service d'Astrophysique, \\CEA Saclay, F-91191 GIF-SUR-YVETTE CEDEX, France.}

\date{}
\maketitle

\pagerange{\pageref{firstpage}--\pageref{lastpage}} \pubyear{2013}

\begin{abstract}
We present GLIMPSE -- Gravitational Lensing Inversion and MaPping with Sparse Estimators -- a new algorithm to generate density reconstructions in three dimensions from photometric weak lensing measurements. This is an extension of earlier work in one dimension aimed at applying compressive sensing theory to the inversion of gravitational lensing measurements to recover 3D density maps. Using the assumption that the density can be represented sparsely in our chosen basis -- 2D transverse wavelets and 1D line of sight dirac functions -- we show that clusters of galaxies can be identified and accurately localised and characterised using this method. Throughout, we use simulated data consistent with the quality currently attainable in large surveys. We present a thorough statistical analysis of the errors and biases in both the redshifts of detected structures and their amplitudes. The GLIMPSE method is able to produce reconstructions at significantly higher resolution than the input data; in this paper we show reconstructions with $6\times$ finer redshift resolution than the shear data. Considering cluster simulations with $0.05\le z_{cl} \le 0.75$ and $3\times 10^{13} h^{-1}M_\odot \le M_{vir} \le 10^{15}h^{-1}M_\odot$, we show that the redshift extent of detected peaks is typically $1-2$ pixels, or $\Delta z \lesssim 0.07$, and that we are able to recover an unbiased estimator of the redshift of a detected cluster by considering many realisations of the noise. We also recover an accurate estimator of the mass, that is largely unbiased when the redshift is known, and whose bias is constrained to $\lesssim 5\%$ in the majority of our simulations when the estimated redshift is taken to be the true redshift. This shows a substantial improvement over earlier 3D inversion methods, which showed redshift smearing with a typical standard deviation of $\sigma \sim 0.2-0.3$, a significant damping of the amplitude of the peaks detected, and a bias in the detected redshift. 
\end{abstract}


\section{Motivation}

Weak lensing -- the distortion of images of distant galaxies by the gravitational potential associated with intervening massive structures -- has established itself as an excellent tool for constraining the distribution and evolution of dark matter in the Universe, and thereby improving constraints on our cosmological model. For this reason, several large surveys have recently undertaken to study this effect at high precision (e.g. CFHTLens, \cite{cfhtlens}; Dark Energy Survey, \cite{des}; Euclid, \cite{euclidredbook}; LSST, \cite{lsst1,lsst2}). 

The primary goal of most lensing surveys is to place constraints on the cosmological model through 2-point (and potentially higher-order $n$-point) statistics of the weak lensing shear field tomographically as a function of redshift. Two-point statistics have been shown to offer excellent constraints on the cosmological model, dark energy equation of state, and modifications to gravity 
\citep[see, for example, the recent CFHTLens results][]{cfht1,cfht2,cfht3}.

However, there is also considerable interest in using weak lensing measurements to generate maps of the dark matter distribution in two and three dimensions; indeed, this is one of the primary science requirements of the upcoming Euclid mission. Such maps are useful for more than just visualisation and detection of structures. Two-point statistics are only optimal for the capture of Gaussian information, while the lensing field consists of both Gaussian and non-Gaussian structures such as clusters of galaxies. Improved constraints on dark matter and dark energy can come from higher order statistics of the lensing field, such as three point (and higher) statistics and peak counts \citep{bernardeau97,TakadaJain2003,TakadaJain2004,ks05,pires09,berge10,pls12}.

Higher-order statistics are typically easier to compute on density maps, or after applying filtering to the shear field \citep[such as the aperture mass statistic,][]{schneider96}. While techniques exist to generate high-fidelity two-dimensional projected density maps, it was not until recently that several linear methods were developed to reconstruct a full three-dimensional density map from photometric weak lensing measurements.

The objective of linear reconstruction algorithms is to find a linear operator that, when applied to the measured shear $\gamma$, yields an estimate of the underlying matter overdensity distribution $\delta$, such that the estimated solution minimises some functional e.g. the variance of the residual between the estimated signal and the true signal. Since the lensing operator is ill-posed, and therefore impossible to invert directly, some regularisation or prior-based constraint must also be applied. 

The most successful linear methods for 3D lensing density mapping have been derived by \citet{sth09} and \cite{vanderplasetal11}. The \citeauthor{sth09} method combines a Wiener filter, encoding a Gaussian prior on both the noise in the data and the estimate of the solution, with an inverse variance filter, using a tuning parameter to switch between these two filters. This method has been successfully applied on simulations and real data from the STAGES survey \citep{stages}, demonstrating that this method is effective in detecting massive clusters of galaxies up to a redshift of $z\sim 0.6$ \citep{sth09,simonetal11}. However, estimation of the redshifts and amplitudes of detected peaks is confounded by the existence of a bias in the location of detected peaks, a broad smearing of the peak along the line of sight due to a radial ``PSF" introduced by the method, and a damping of the overall amplitude relative to the true density. There is also limited sensitivity at high redshift, an effect which may be improved by including information about the positions of the galaxies for which we have a shear measurement to help constrain the reconstruction \citep{simon12}.

\citet{vanderplasetal11} adopt a different approach: they use as their estimator a simple inverse variance filter, but in order to regularise the inversion they use a singular value decomposition and retain only the largest singular values. This method has been tested on simulated data, and shows comparable results to the method of \citet{sth09}, with similar bias, smearing and damping seen in the reconstructed density maps.

In this paper, we present GLIMPSE -- Gravitational Lensing Inversion and MaPping using Sparse Estimators -- a new, sparsity-based approach to weak lensing reconstructions in 3D. Building on the earlier work by \cite{LDS12} (hereafter LDS12), this inherently non-linear method uses the prior assumption that the density can be sparsely represented in an appropriate basis set (dictionary) in order to regularise the inversion of the lensing operator. The simple, one-dimensional implementation of this idea presented in LDS12 showed substantial improvement over the earlier linear methods. We now present a full three-dimensional sparse inversion method using an improved algorithm, and demonstrate it to be effective and robust at detecting clusters of galaxies, and constraining both their mass and redshift.

\section{Weak lensing in 3D}
\label{sec:3dlensing}
\subsection{Theory}

Weak lensing describes the distortion of the images of distant galaxies arising from the perturbation of their light's path by the presence of massive haloes along the intervening line of sight. As such, gravitational lensing represents an integrated effect along a given line of sight. This can be related to an effective lensing \textit{convergence}, $\kappa$:
\small
\begin{eqnarray}
  \kappa(\bt,w) = \frac{3H_0^2\Omega_M}{2c^2} \int_0^w dw^\prime \frac{f_K(w^\prime) f_K(w-w^\prime)}{f_K(w)}\frac{\delta[f_K(w^\prime)\bt,w^\prime]}{a(w^\prime)}\ ,\nonumber\\ 
  \label{eq:kapconv}
\end{eqnarray}\normalsize
where $\delta \equiv \rho(\boldsymbol{r})/\overline{\rho} - 1$ is the matter overdensity, $H_0$ is the hubble parameter, $\Omega_M$ is the matter density parameter, $c$ is the speed of light, $a(w)$ is the scale parameter evaluated at comoving distance $w$, and 
\begin{equation}
  f_K(w) = \begin{cases}  K^{-1/2}\sin(K^{1/2}w), & K>0 \\ w, & K=0 \\ (-K)^{-1/2}{\rm sinh}([-K]^{1/2}w) & K<0 \end{cases}\ ,
\end{equation}
gives the comoving angular diameter distance as a function of the comoving distance and the curvature, $K$, of the Universe.

The convergence, $\kappa$, is therefore a projected, dimensionless surface density, and can be related to the gravitational \textit{shear}, $\gamma$, via a 2D convolution
\begin{equation}
	\gamma(\bt) = \frac{1}{\pi}\int\ d^2\bt^\prime \mathcal{D}(\bt - \bt^\prime)\kappa(\bt^\prime)\ ,
	\label{eq:gamconv}
\end{equation}
where 
\begin{equation}
	\mathcal{D}(\bt) = \frac{1}{(\bt^\ast)^2} ,
\end{equation}
$\bt = \theta_1+{\rm i}\theta_2$ is an angular position expressed in complex notation, and an asterisk $^\ast$ represents complex conjugation. 

The measured shear is thus related to the matter overdensity by a 2D transverse convolution and a 1D line of sight convolution, and can be expressed simply in matrix notation as
\begin{equation}
\gamma = \mathbf{P}_{\gamma\kappa}\mathbf{Q}\delta\ +\varepsilon,\ \ \ \ \ \varepsilon \sim \mathcal{N}(0,\sigma^2)\ ,\label{eq:gamdelta}
\end{equation}
where $\mathbf{P}_{\gamma\kappa}$ represents the transverse convolution in equation \eqref{eq:gamconv}, $\mathbf{Q}$ represents the line of sight convolution in equation \eqref{eq:kapconv}, and $\varepsilon$ represents measurement noise, typically assumed to be white Gaussian shape noise.

%\subsection{Linear reconstruction algorithms}

\subsection{Sparsity-based methods}

Using a sparsity prior to regularise a linear inverse problem has proven very successful in a large variety of domains, and in particular in astronomy and cosmology \citep[see][and references therein]{fs09,smf10}. Let us consider the following general linear inverse problem
\begin{equation}
y = \mathbf{A} x + n\ ,
\end{equation}
where $x \in \mathbb{R}^N$ is the signal we aim to recover, $y \in \mathbb{R}^m$ are the noisy measurements and $n$ is an additive noise of bounded variance $\sigma_n^2$. $\mathbf{A}$ is a bounded linear operator, which is typically ill-posed, and in the general case, $m$ can be smaller than $N$. 

To regularise this problem, the signal is assumed to be sparsely represented in an appropriate overcomplete dictionary $\mathbf{\Phi}$: $x = \mathbf{\Phi} \alpha$, i.e. only a few non zero coefficients $\alpha$ are necessary to express the signal $x$ in the dictionary $\Phi$. In practice, for most problems the signals of interest are not strictly sparse, but this assumption can be relaxed to compressible signals; when expressed in the appropriate dictionary (Fourier, Wavelets, Curvelets, etc.) the coefficients of a compressible signal exhibit a polynomial decrease of their sorted absolute values. 

Under this assumption, the linear inverse problem can be cast as a convex optimization problem of the form
\begin{equation}
\centering
\min_{\alpha} \frac{1}{2} \parallel y - \mathbf{A} \mathbf{\Phi} \alpha \parallel^{2}_{2} + \lambda \parallel \alpha \parallel_{1}\ .
\label{eq:Lagrangian}
\end{equation}
This formulation is known as the augmented Lagrangian form of the Basis Pursuit DeNoise (BPDN) problem. The first term in this expression imposes a quadratic constraint on the data fidelity of the reconstruction whereas the second term uses the sparsity promoting $\ell_1$ norm to impose a sparse reconstruction. The parameter $\lambda$ controls the sparsity of the solution.

A first non-linear sparsity-based approach to the 3D weak lensing density mapping has been presented in LDS12 under the framework of compressed sensing. This work reduced the 3D weak lensing reconstruction problem to one dimension, considering the reconstruction of each line of sight independently. Despite the loss of information this implies, the method was seen to significantly reduce the biasing, smearing and damping effects seen in linear reconstruction methods. In addition, LDS12 demonstrated that this approach permits reconstruction of the density contrast at a higher resolution than that of the input data, and showed that this method is effective both at detecting structures at high redshift ($z\sim1$), and in disentangling the lensing signal when there are multiple structures along the line of sight. 

In this work, we seek to extend the method of LDS12 to include a fully three-dimensional treatment of the inversion problem, using as our data the measured shear $\gamma$, rather than the line-of-sight convergence, $\kappa$, in LDS12. A drawback of the method proposed in LDS12 is that it included a tuning parameter, $\epsilon$, responsible for noise control in the algorithm. This parameter was free to be chosen by the user, with little physical motivation or strategy presented for choosing an appropriate value. In this paper, we will present an algorithm whose noise control threshold is related in a straightforward way to the noise in the data, thereby giving the user a simple, intuitive strategy for choosing an appropriate threshold.

\section{The GLIMPSE method}
\label{sec:walrus}

\subsection{From lensing tomography to 3D dark matter density}

The problem we seek to invert has been introduced in equation (\ref{eq:gamdelta}) in its exact continuous form. Following the approach of \cite{sth09}, we introduce a discrete version of the lensing operator using tomographic redshift binning.

We consider that we have a catalog of measured ellipticities $\mathbf{E}_k$ from galaxies at angular positions $\bt_k$ and photometric redshifts $z_k$. For the purpose of reconstructing the 3D density contrast, this catalog can be binned into $N_z$ tomographic shear maps $\bg^{1}, \ldots,\bg^{n}$ of size $N_x \times N_y$ where $N_x$ and $N_y$ are the number of pixels on the 2D cartesian grid in the angular domain. On each pixel, the shear value is estimated by averaging the ellipticities of the galaxies belonging to this pixel:
\begin{equation}
\bg_{i,j}^{n} = \frac{\sum\limits_{k \in \mathbb{G}_{i,j}^n} w_k \mathbf{E}_k}{\sum\limits_{k \in \mathbb{G}_{i,j}^n} w_k }\ ,
\end{equation}
where $w_k$ is a statistical weight for the measurement on galaxy $k$ and $\mathbb{G}_{i,j}^n$ contains the indices of the galaxies belonging to the pixel $(i,j)$ of the redshift bin $n$.

\subsection{Redshift binning and lensing operator}
\label{sec:zbins}
The lensing operator in Eq. (\ref{eq:gamdelta}) relates the density contrast to the convergence and shear at a given comoving distance. The $\mathbf{P}_{\gamma\kappa}$ operator can be straightforwardly computed in Fourier space, but the line-of-sight operator $\mathbf{Q}$ must be discretised, and needs to take into account the redshift distribution of the galaxies in our measurement catalogue. 

Previous 3D weak lensing works have all chosen to bin their shear catalogues in redshift bins that are of equal width $\delta z$. We chose to adopt a different approach, and to bin our shear maps adaptively such that each redshift bin contains the same mean number of galaxies. This is preferable and more intuitive, as it ensures that the shape noise level is the same on each redshift slice.

We assume that our sources are distributed in redshift according to some probability distribution function $p(z)$. Given the redshift distribution of sources in bin $n$, the convergence $\kappa_{i,j}^n$ in this bin is related to the density contrast through
\begin{equation}
\kappa^n_{i,j}  =  \frac{3H_0^2\Omega_M}{2c^2} \int_0^{w_h} dw\ g^n(w) f_K(w) \frac{\delta(f_K(w)\bt_{i,j})}{ a (w)}\ ,
\label{eq:kapdelta}
\end{equation}
with
\begin{equation}
g^{n}(w) = \int_w^{w_h} dw^\prime \frac{f_K(w^\prime - w)}{f_K(w^\prime)} \left\lbrace p^n(z)\frac{dz}{dw} \right\rbrace_{z=z(w^\prime)}\ .
\end{equation}

To build a discrete lensing efficiency operator, we bin the overdensity distribution into $N_{lp}$ lens planes $\delta^n$. Equation (\ref{eq:kapdelta}) can therefore be approximated by a discrete sum over lens planes along a given line of sight:
\begin{equation}
	\kappa^{n}_{i,j} = \sum\limits_{p=1}^{N_{lp}} \mathbf{Q}_{n p} \delta^p_{i,j}\ ,
\end{equation}
where $\mathbf{Q}$ is now a matrix operator defined by the following matrix elements:
\begin{equation}
	\mathbf{Q}_{n p} = \frac{3H_0^2\Omega_M}{2c^2} \int_{w_p}^{w_{p+1}} dw\ g^n(w)  \frac{f_K(w)}{ a (w)}\ .
\end{equation}

\subsection{Choice of an adapted dictionary}

	Many experiments, in particular cosmological N-body simulations, have shown the dark matter to be largely distributed in halos connected by thin filaments. We limit ourselves in this paper to the detection of clusters of galaxies; however we note that the method presented here is entirely general, and other dictionaries can be added to our algorithm to detect other types of structure with no modification to the overall approach (e.g. ridgelets or curvelets to detect filamentary structures).
	
At the redshift resolution we are able to attain, clusters of galaxies are small compared to the radial length of a given redshift bin. Therefore, a dark matter halo can be considered to have no radial depth and can be represented as a flat disc. This prompts us to choose a 2D-1D dictionary $\mathbf{\Phi}$ composed of Isotropic Undecimated Wavelets in the 2D angular plane and Dirac $\delta$-functions along the radial dimension. The wavelet transform used in this work is the Starlet introduced in \citet{Starck07Starlet}, which is particularly well suited to represent positive, isotropic objects.
	 
\begin{figure}
\centering
	 \includegraphics[width=0.45\textwidth]{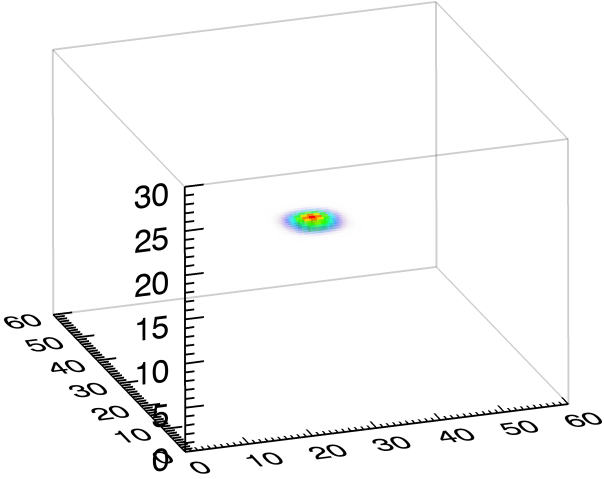}
	 \caption{Representation of single atom of our 2D-1D dictionary composed of Isotropic Undecimated Wavelet in the angular domain and Diracs in the radial domain}
	 \label{fig:2D1Datom}
\end{figure}

\subsection{Reconstruction algorithm}

The GLIMPSE algorithm aims to solve a BPDN problem of the form in equation \eqref{eq:Lagrangian}. Using the specific notations of the weak lensing reconstruction problem, the optimization problem can be written as:
\begin{equation}
\centering
\min_{\alpha} \frac{1}{2} \parallel \mathbf{\Sigma}^{-1/2} \left[ \gamma - \mathbf{P}_{\gamma\kappa} \mathbf{Q} \mathbf{\Phi} \alpha  \right] \parallel^{2}_{2} + \lambda \parallel \alpha \parallel_{1}\ ,
\label{eq:WLLagrangian}
\end{equation}
where $\alpha$ are the coefficients of the density contrast $\delta$ in the dictionary  $\mathbf{\Phi}$ introduced in the previous section and $\mathbf{\Sigma}$ is the diagonal covariance matrix of the shear measurements. 

Many algorithms have been proposed to address this kind of optimisation problem, imposing the $\ell_1$ sparsity constraint through soft thresholding. In this work, we base our reconstruction algorithm on the Fast Iterative Soft Thresholding Algorithm (FISTA) presented in \citet{FISTA2009}. This algorithm is a fast variant of the standard Iterative Soft Thresholding (ISTA), which is widely used in sparse linear inversion. In order to use this algorithm to solve problem (\ref{eq:WLLagrangian}), we first apply a change of variable to normalise the columns of the $\mathbf{Q}$ matrix. This step is necessary to ensure a good behaviour of this gradient descent based algorithm, and is achieved by replacing the operator $\mathbf{Q}$ by $\mathbf{Q^\prime} = \mathbf{Q} {\mathcal N}^{-1}$ where ${\mathcal N}$ is a diagonal matrix  with diagonal elements equal to the square  of the $\ell_2$ norm of the corresponding column of the matrix $\mathbf{Q}$:
	\begin{equation}
	\forall 1 \leq n \leq N_{lp},\quad	{\mathcal N}_{nn} = \sum_p \mathbf{Q}_{p n}^2\ . 
	\label{eq:normalisation}
	\end{equation}
This change of operator corresponds to the change of variable $\delta \rightarrow \delta^\prime$ (and equivalently $\alpha \rightarrow \alpha^\prime$) defined by ${\delta^\prime}_{i,j}^n = \mathcal{N}_{n n} \delta_{i,j}^n$.

With this change of variable, the basic iteration for both FISTA and ISTA algorithms is
\begin{equation}
X_{n+1} =  ST_{\lambda} \left( \alpha_{n}^\prime + \mu \mathbf{\Phi}^t  \mathbf{Q^\prime}^t \mathbf{P}_{\gamma\kappa}^t \mathbf{\Sigma}^{-1} \left[\gamma -  \mathbf{P}_{\gamma\kappa} \mathbf{Q^\prime} \mathbf{\Phi} \alpha_{n}^\prime \right] \right)\ ,
\label{eq:update}
\end{equation}
where $ST_{\lambda}$ is the soft thresholding operator
\begin{equation}
ST_{\lambda}(x) = {\rm sgn}(x)\max(|x| - \lambda,0)
\end{equation}
and $\mu$ is the gradient descent step ensuring convergence of the algorithm. $X_{n+1}$ is an updated estimation of the coefficients $\alpha^\prime$ of the reconstruction. The particularity of the FISTA algorithm is that the actual update of the coefficients requires an additional step amounting to a simple weighted average between current and previous estimates:
\begin{eqnarray}
t_n & = & \frac{1 + \sqrt{1 + 4t_n^2}}{2}\ , \\ 
\alpha_{n+1}^\prime & = & X_{n+1} + \left( \frac{t_n -1}{t_{n+1}} (X_{n+1} - X_{n})  \right)\ ,
\end{eqnarray}
with $t_0=1$ and $X_0={\alpha^\prime}_0$. 

This algorithm converges as long as the gradient descent step $\mu$ verifies
\begin{equation}
	0 < \mu < \frac{1}{\parallel \mathbf{\Phi}^t \mathbf{Q^\prime}^t \mathbf{P}_{\gamma\kappa}^t \mathbf{\Sigma}^{-1} \mathbf{P}_{\gamma\kappa} \mathbf{Q^\prime} \mathbf{\Phi}\parallel } \ ,
	\label{eq:mu}
\end{equation}
where $\parallel \cdot \parallel$ is the spectral norm of the operator. 
The solution provided by this algorithm is the normalised density contrast ${\delta}^\prime$ from which one recovers the unnormalised density contrast by applying the ${\mathcal N}^{-1}$ matrix along the lines of sight.

GLIMPSE is an implementation of FISTA which differs only in the thresholding operator used in equation \eqref{eq:update}. Standard soft thresholding is known to introduce a bias in the reconstruction, as soft thresholding decreases the amplitude of all of the coefficients in the reconstruction. To circumvent this issue, we use firm thresholding \citep{Gao97}, rather than standard soft thresholding. Firm thresholding results in an unbiased solution by defining a significance criterion above which coefficients will no longer be altered or shrunk. Sparsity is imposed by shrinking all coefficients below the required significance level, with a shrinkage that is dependent on the amplitude of the coefficient.

The firm thresholding operator therefore depends on two parameters, $\lambda_1$ and $\lambda_2$, and is defined by
\begin{equation}
	\quad FT_{\lambda_1,\lambda_2}(x) = \left\lbrace \begin{matrix} 
	0, & \mbox{if } |x| \leq \lambda_1,\\
	{\rm sgn}(x) \frac{\lambda_2 (| x| - \lambda_1)}{\lambda_2 - \lambda_1}, & \mbox{if } \lambda_1 < |x| \leq \lambda_2,\\
	x,  &  \mbox{if } | x| > \lambda_2.
\end{matrix} \right.
\label{eq:firmthresh}
\end{equation}
The parameter $\lambda_1$ sets the minimum significance level above which coefficients are retained in the solution estimate, thereby effectively denoising the solution, while $\lambda_2$ defines the significance level above which the shrinkage vanishes. 

In order to provide an intuitive method for choosing $\lambda_1$, we relate it directly to the noise level $\sigma^p$ in redshift bin $p$, estimated at each iteration by considering the MAD of the residual
\begin{equation}
	\sigma^p = \frac{\mu}{0.6747} \MAD \left( \left[\mathbf{Q^\prime}^t \mathbf{P}_{\gamma\kappa}^t \mathbf{\Sigma}^{-1} \left[\gamma -  \mathbf{P}_{\gamma\kappa} \mathbf{Q^\prime} \mathbf{\Phi} \alpha_{n}^\prime \right] \right]^p_{i,j} \right).
\end{equation}
With this estimation of the noise level, the threshold $\lambda_1$ is set for each redshift plane in the reconstruction to $\lambda_1 = k \sigma^p$, where the level of significance $k$ is iteratively adjusted using a decreasing threshold scheme until reaching a minimum level $k_{min}$ set by the user. For the reconstructions presented here, we choose $k_{min} = 4.5$. We further choose $\lambda_2 = \sigma^p/\mu$, such that reconstructed coefficients with this significance level are not affected by shrinkage. This results in an unbiased estimator of the density contrast.

At the first iteration, the most significant wavelet coefficient (highest SNR) is selected and the level $k$ is set to the average between the first and second highest SNR coefficients in the residuals. Therefore, at the beginning only this selected coefficient is allowed to enter the reconstruction. An update of the level $k$ occurs whenever the SNR $s$ of a different wavelet coefficient becomes greater than the SNR $s_0$ of this selected coefficient. This update is controlled by a sufficient decrease parameter $c_{sd}$, which we set to $0.1$, and is only allowed when $ \frac{s - s_0}{s - k_{min}} \geq c_{sd}$. When this condition is met, the most significant wavelet coefficient is selected and the threshold level is decreased to the average between the first and second highest SNR in the residuals. The procedure is repeated until the minimum thresholding level $k_{min}$ is reached. This will lead to a slowly decreasing threshold, allowing at most only one coefficient to become active at each iteration. This scheme further promotes the sparsity of the solution and improves the speed of the reconstruction.

%For $\lambda_2$, we use a fixed SNR level for each redshift plane $p$ $\lambda_2 = \frac{\sigma^p}{\mu}$. This choice of parameter ensures that the reconstructed amplitude of any structure with a SNR in density contrast higher than $1/\mu$ will not be biased by the thresholding.

Finally, we do not want to impose the same sparsity constraint along the line of sight and in the tangential plane. To relax the sparsity constraint on each redshift plane, we identify the real space support of detected wavelet coefficients and we set the thresholding level $k$ for any other wavelet coefficients within this spatial domain to the lowest level $k_{min}$.

The full description of our method is provided in Algorithm 1.

\begin{algorithm}
 \caption{GLIMPSE reconstruction algorithm}
 \SetAlgoLined
 \KwData{
 	\begin{itemize}
 		\item $\gamma$: Complex 3D array of the binned shear
 		
 		\item $\Sigma$: 3D array of the covariance in each voxel
 	\end{itemize} }
 \KwResult{$\delta$: 3D array of the reconstructed density contrast} 
 
 \KwParameters{\begin{itemize}
 	\item Number of wavelet scales $N_{scale}$
 	
 	\item Minimum threshold level $k_{min}$ 
 	
 	\item Sufficient decrease parameter $c_{sd}$
\end{itemize}  }

 \KwInitialization{  $\alpha_0^\prime = \mathbf{0}$, $t_0=1$, $X_0 = \alpha_0^\prime$, $s_0 = 0$, $(x_0,y_0,z_0,j_0) = (0,0,0,0)$\\ $\mu $ set according to equation (\ref{eq:mu})}
 
 \bigskip
 
 \For{$n=0 \mbox{ to } N_{iter} -1$}{

 \tcc{Estimated density contrast}
  $\delta_n^\prime = \Phi \alpha_n^\prime$
 \bigskip
 
 \tcc{Forward gradient descent step}
  $r_n = \mu \mathbf{\Phi}^t  \mathbf{Q^\prime}^t \mathbf{P}_{\gamma\kappa}^t \mathbf{\Sigma}^{-1} ( \gamma -   \mathbf{P}_{\gamma\kappa} \mathbf{Q^\prime}  \delta_n^\prime) $
  $\tilde{X}_{n+1} = \alpha_n +  r_n$
 \bigskip
 
 \tcc{Noise estimation on each redshift bin and each wavelet scale}
 \For{$p=1 \mbox{ to } N_{lp}$ and $j = 1 \mbox{ to } N_{scale}$}{
	 $\sigma^p_j = \MAD(r^p_{n,j}) / 0.6747$
 }
 
 \bigskip
  
 \tcc{Update of the threshold level\\ $\max_2$ gives the second highest value}
 $s_0 = |r_{n,j_0}^{z_0}[x_0,y_0]|/\sigma^{z_0}_{j_0}$
 
 $s = \max(|r_n| / \sigma) $
 
 \If{$ s - s_0 \geq c_{sd}( s - k_{min} ) $}{ 
 	
 	$k = ( s + \max_2(|r_n| / \sigma) )/2$  
 	
 	$(x_0,y_0,z_0,j_0) \leftarrow$ set to indices of maximum SNR coefficient
 }
 \bigskip
  
 \tcc{Backward thresholding step}
 \ForAll{ $x,y,p$ and $j = 1 \mbox{ to } N_{scale}$ }{ 
 		\eIf{${\delta^\prime}_n^p[x,y] \neq 0$}{
 			$X_{n+1,j}^p[x,y] = FT_{k_{min} \sigma^p_j,\sigma^p_j/\mu}( \tilde{X}_{n+1,j}^p[x,y])$
		}{
			$X_{n+1,j}^p[x,y] = FT_{k \sigma^p_j,\sigma^p_j/\mu}( \tilde{X}_{n+1,j}^p[x,y])$
		}	
	
 }
 
 \bigskip 
 
 \tcc{FISTA specific update of the coefficients}
 $t_n = \frac{1 + \sqrt{1 + 4t_n^2}}{2}$  
 $\alpha_{n+1}^\prime  =  X_{n+1} + \left( \frac{t_n -1}{t_{n+1}} (X_{n+1} - X_{n})  \right)$
 }
 \tcc{Recover density contrast}
 $\delta = N \mathbf{\Phi} \alpha_{N_{iter}}^\prime $
\end{algorithm}

\section{Cluster simulations}
\label{sec:sims}

The GLIMPSE method is nonlinear, and therefore the performance of the algorithm cannot be derived analytically, but rather needs to be assessed through numerical simulations. To do this, we generated a suite of cluster simulations, spanning a range of redshifts between $0.05\le z\le 0.75$ and virial masses in the range $3\times10^{13}h^{-1}M_\odot \le M_{\vir} \le 10^{15}h^{-1}M_\odot$. For the density profile of the clusters we adopted an NFW \citep{NFW1997} profile given by
\begin{equation}
	\rho(r) = \frac{\rho_s}{(c r/r_{\vir}) (1 + c r/r_{\vir})^2}\ ,
\end{equation}
where $\rho_s$ is the central density parameter, $r_{\vir}$ is the virial radius and $c$ is the concentration parameter. As described in \cite{TakadaJain2003}, the number of degrees of freedom for this profile can be reduced to 2, keeping only a dependence in mass $M$ and redshift $z$. The central density parameter $\rho_s$ can be eliminated from the definition of the virial mass
\begin{equation}
M = \int_0^{r_{\vir}} 4 \pi r^2 dr \rho(r) = \frac{ 4 \pi \rho_s r_{\vir}^3}{c^3} \left[ \ln(1+c) - \frac{c}{1+c} \right]\ .
\end{equation} 
Then the virial radius can be linked to the mass of the halo through the spherical collapse model
\begin{equation}
M = \frac{4 \pi}{3} \overline{\rho}_0 \Delta_{\vir}(z)  r_{\vir}^3\ ,
\end{equation}
where $\overline{\rho}_0$ is the mean density of matter today and $\Delta_{\vir}(z)$ is the critical overdensity of collapse at redshift $z$. Following \cite{couponetal12}, we use the fitting formula for $\Delta_{\vir}(z)$ from \cite{Weinberg2003}:
\begin{equation}
\Delta_{\vir}(z) = 18 \pi^2 \left[ 1 + 0.399 (\Omega_m^{-1} - 1 )^{0.941} \right]\ .
\end{equation}
Finally, we assume a mass and redshift dependance for the concentration parameter $c(M,z)$ given by
\begin{equation}
c(M,z) = c_0 (1 + z)^{-1} \left[ \frac{M}{M_{\star}} \right]^\beta\ ,
\end{equation}  
where $M_{\star}$ is the non-linear mass scale at present day defined by $\delta_c(z=0) = \sigma(M_{\star})$, in which $\delta_c$ is the linear critical density and $\sigma(M)$ is the RMS of density fluctuations in a sphere of radius $(3/4 \pi M \overline{\rho}_0)^{1/3}$. We adopt the parameterisation from \cite{couponetal12}: $c_0 = 11$ and $\beta = 0.13$.

These relations allow us to parameterise the simulated halos only by their mass and redshift within the framework of the halo model. From the NFW density profile, we computed the corresponding shear signal, which is derived analytically in \cite{TakadaJain2003}. The computations required to simulate the halos were performed making extensive use of the NICAEA software package\footnote{http://www2.iap.fr/users/kilbinge/nicaea/}, using a flat $\Lambda$CDM cosmology with $\Omega_{\rm M} = 0.264,\ \Omega_\Lambda = 0.736$, and $H_0 = 71$km/s/Mpc. 

The shear signal was simulated for each cluster in an otherwise empty field of $1^\circ \times 1^\circ$, with an angular pixel size of $1^\prime\times 1^\prime$ on 30 tomographic redshift bins, with a galaxy redshift distribution given by
\begin{equation}
	n(z) = z^{\alpha} \exp\left( - \left[\frac{z}{z_0}\right]^{\beta}  \right)\ ,
\end{equation}
where we take $z_0 = 1/1.4$, $\alpha = 2$ and $\beta = 1.5$. 

\begin{figure}
\centering
\includegraphics[width=0.45\textwidth]{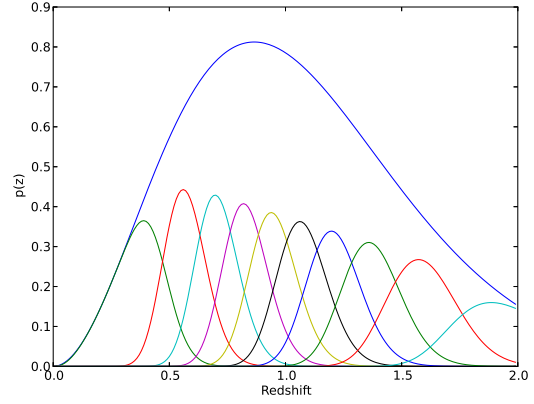}
\caption{Global redshift probability distribution and 10 photometric redshift bins with an equal number of galaxies per bin assuming unbiased photometric redshift estimates with Gaussian errors whose width varies as a function of redshift as $\sigma = 0.05(1+z)$.}
\label{fig:pz}
\end{figure}

In this work, we consider the redshift information to be provided by photometric redshift measurements with Gaussian errors with a standard deviation that varies as a function of redshift as $\sigma_z (1+z)$, and a potential bias $z_{bias}$, but without catastrophic failures. Following \cite{maetal06}, the true redshift distribution $p^n(z)$ of a tomographic bin $n$ can be obtained through:
\begin{equation}
p^n(z) = \frac{1}{2} p(z) \left[ \erf(x_{n+1}) - \erf(x_{n}) \right]
\end{equation}
with
\begin{equation}
x_n = \frac{z_{n+1} - z + z_{bias}}{\sqrt{2} \sigma_z (1 + z) }.
\end{equation}
For the work presented here, we took $z_{bias} = 0$ and $\sigma_z = 0.05$, the latter being in line with the minimum required accuracy for photometric redshifts in the Euclid survey. The global redshift distribution is plotted in figure \ref{fig:pz}, with the galaxy distribution for 10 tomographic redshift bins overplotted, each containing the same number of galaxies per bin.

\begin{figure}
\centering
\includegraphics[width=0.45\textwidth]{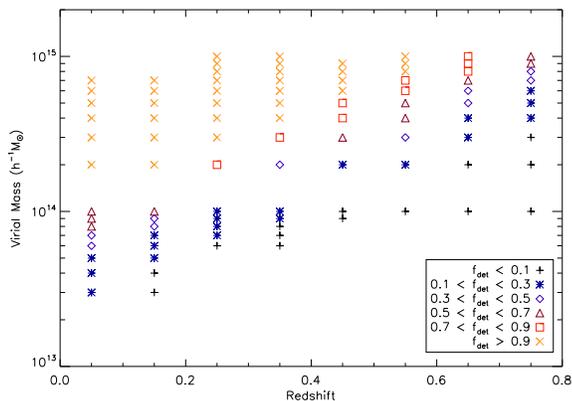}
\caption{Mass and redshift distribution of simulated cluster haloes. The different coloured symbols show the fraction of noise realisations $f_{det}$ for which a given cluster was detected using GLIMPSE out of the 1000 different noise realisations of the cluster field.}
\label{fg:samplepoints}
\end{figure}

The final simulated fields were obtained by adding independent Gaussian shape noise to the shear signal with a standard deviation of $\sigma_\varepsilon = 0.25$, and we assumed a number density of galaxies of $n_g = 30$/arcmin$^2$. For each halo, we generated 1000 such noisy fields from independent noise realisations, in order to estimate our reconstruction errors. Figure \ref{fg:samplepoints} shows the distribution in virial mass and redshift of the 96 haloes simulated. The values were chosen to approximately trace the mass function (at the high mass end) and our estimated detection limits (at the low-mass end), but these sample points should not be taken to represent a complete sample of haloes detectable with our method. The different colours and symbols in the figure show the fraction of noise realisations $f_{det}$ for which a given cluster was detected using GLIMPSE.

Note that, for these simulations, we assumed pixellated data from the start, thereby implicitly assuming a uniform distribution of galaxies in $x$ and $y$. This is not realistic; however, the method presented is entirely general, and can account for binned/pixellated data where the noise level per pixel varies. In addition, an extension of the method is currently under development which will be able to work directly on a galaxy catalogue, without the need for pixellation of the shear data. This will be presented in a future work. 
 
%The density contrast maps generated, and which we aim to reconstruct, consist of 60 redshift bins between $z=0$ and $z=2$. This means that the radial pixel length is large compared to the typical size of a cluster, and each cluster can be considered to lie completely within a single redshift plane. 

%\subsection{N-body simulations}

%In addition to individual cluster simulations, we have also generated a $4 \times 4$ square degree lensing field from an N-body simulation. .... \textbf{[add more details on the simulation itself ?]}

%From the lightcone, we use the CALCLENS\footnote{http://code.google.com/p/calclens} software described in \cite{Becker2012}. CALCLENS allows us to generate 150 shear maps at regular interval in comoving distance between $z=0$ and $z=2$. Using the same parameters as for the individual cluster simulations, we use these maps to generate the lensing signal for each of 30 tomographic redshift bins. Again, we produce a thousand independent noise realizations for this field.

\section{Algorithm Performance}
\label{sec:results}
\begin{figure*}
\begin{center}
\begin{tabular}{c c c}
\subfigure[Input density contrast]{\includegraphics[width=0.3\textwidth,trim=10mm 30mm 10mm 30mm, clip]{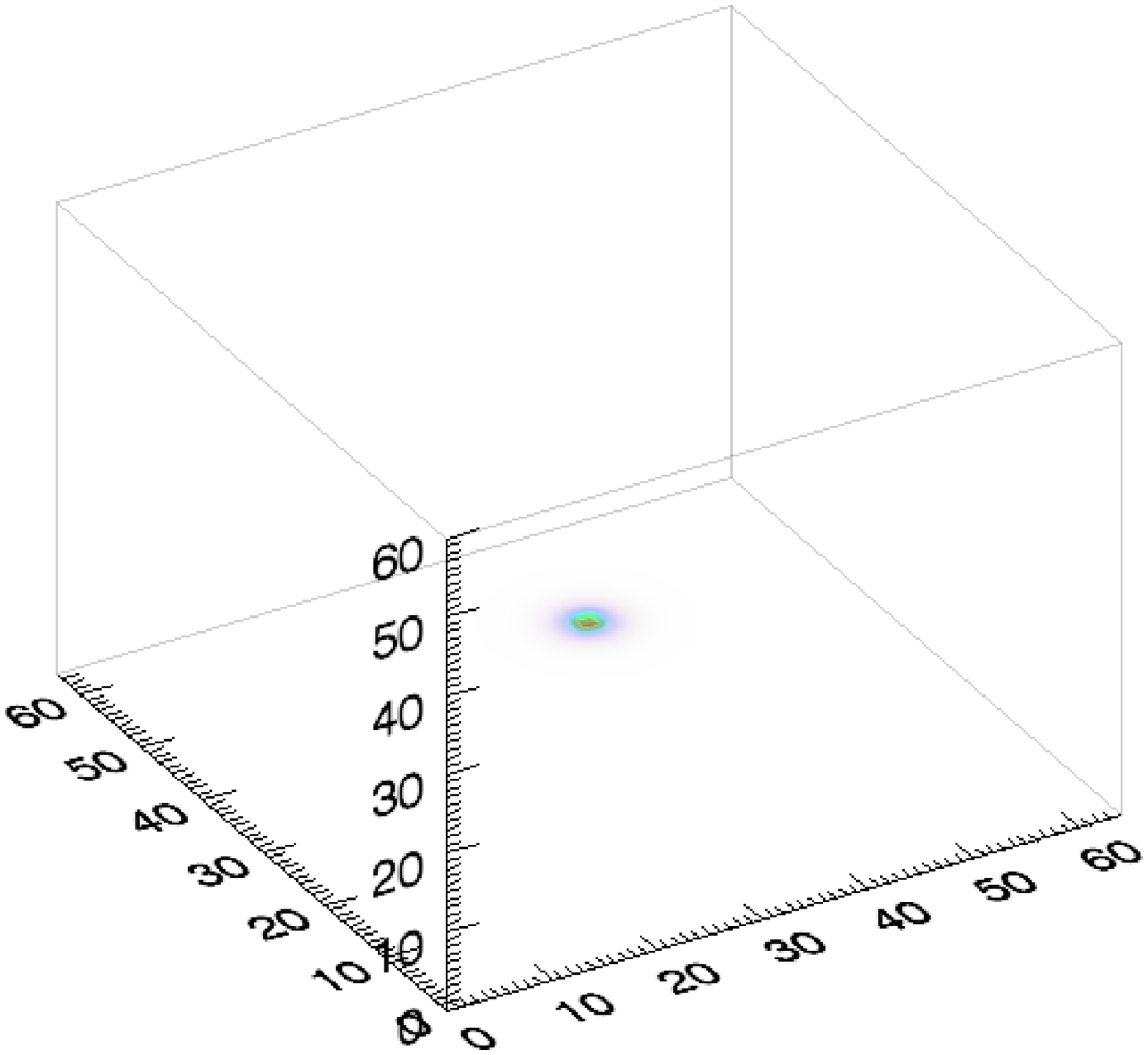}} & 
\subfigure[Reconstructed density contrast]{\includegraphics[width=0.3\textwidth,trim=10mm 30mm 10mm 30mm, clip]{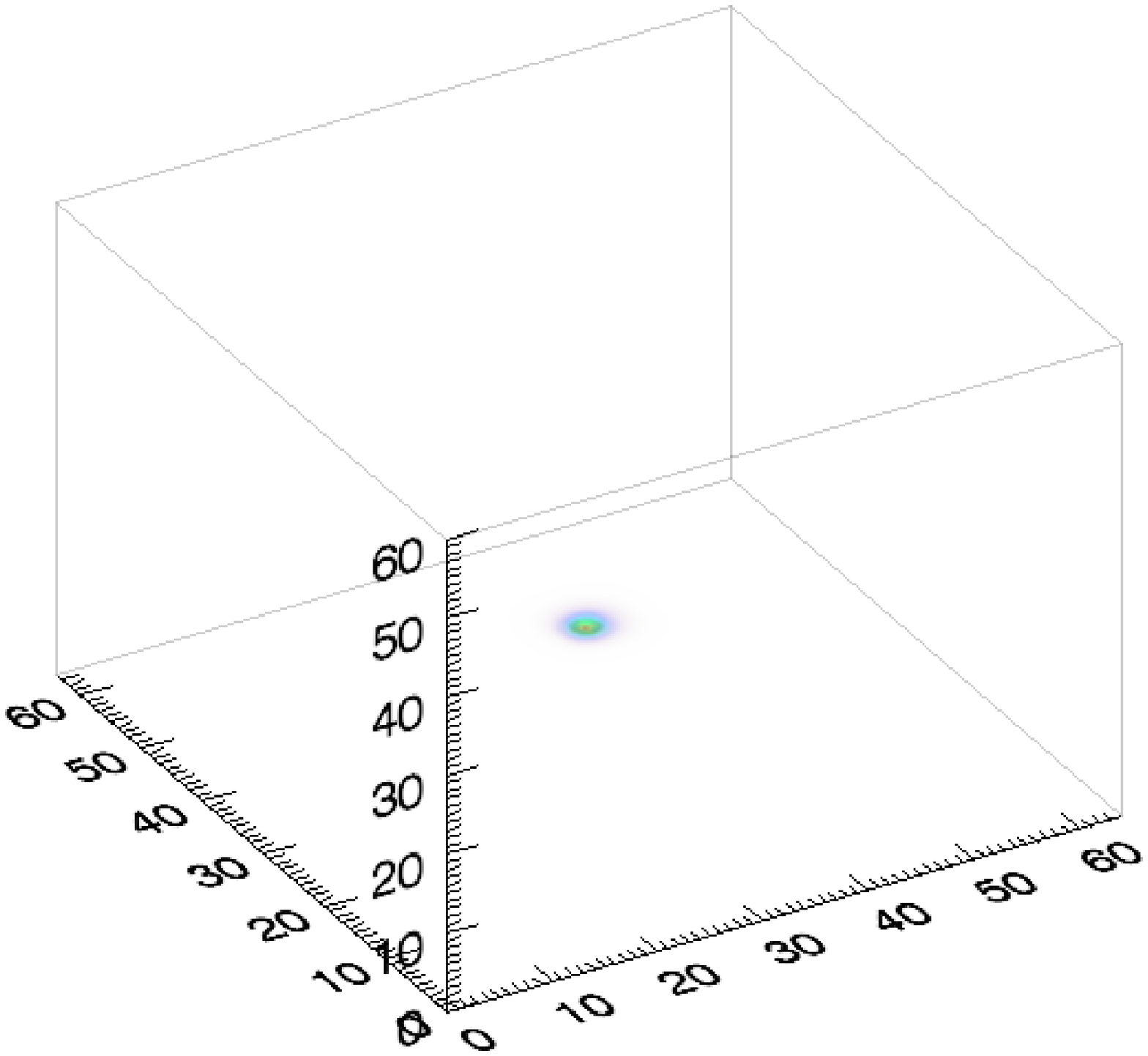}} &
\subfigure[Mean density contrast along the central four lines of sight of the cluster from the input (black diamond) and the reconstruction (orange triangle)]{\includegraphics[width=0.3\textwidth,trim=10mm 40mm 20mm 40mm, clip]{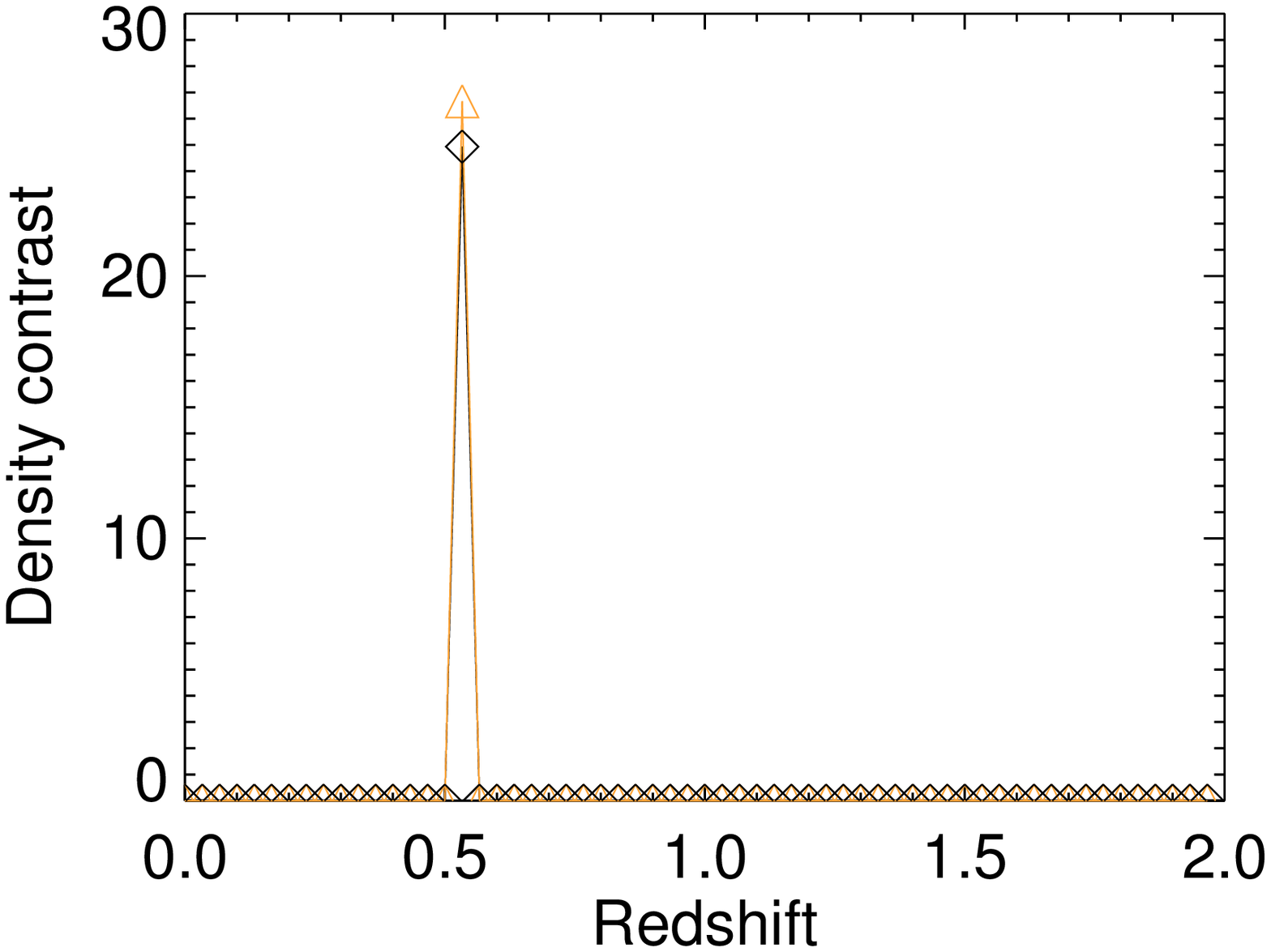}}
\end{tabular}
\vskip-2ex
\caption{Typical GLIMPSE reconstruction of an NFW halo of mass $10^{15} h^{-1} M_{\odot}$ at redshift $z=0.55$. $n_{g}$ = 30 arcmin$^{-2}$ and $\sigma_\epsilon = 0.25$. The $x$ and $y$ axes in panels (a) and (b) are in arcminutes, while the $z$ axis is in pixels, with $\Delta z_{pix} = 0.033$.}
\label{Wavelet_decomposition}
\end{center}
\end{figure*}

For each cluster in our simulations suite, we now have 1000 reconstructions carried out under 1000 different noise realisations. An example of one such simulation and its reconstruction is shown in figure \ref{Wavelet_decomposition}. 

In order to identify reconstructed peaks in the reconstructions, we used the Clumpfind algorithm\footnote{http://www.ifa.hawaii.edu/users/jpw/clumpfind.shtml} \citep{williamsetal94}. This method identifies connected pixels above a given threshold within a 3D map, and is able to effectively deblend overlapping structures. We set $\delta_{\rm min} = 1$ to be the minimum threshold for inclusion as part of a detected peak. For each detected peak, we computed the $\delta-$weighted centroid, and defined this to be the detected position of the peak. 

\subsection{False detections}

Occasionally, spurious peaks not associated with the lensing cluster will appear in the reconstruction as a result of noise in the data. Indeed, this was one of the major problems with the LDS12 approach, whereby numerous high-redshift false detections were seen in their cluster reconstructions. 

The incidence of such false peaks is controlled by the threshold $\lambda_1$ chosen in the GLIMPSE algorithm. In the results presented here, this threshold was chosen to be at $4.5 \sigma$. As these spurious peaks arise out of the noise, one would not expect these detections to persist when considering a different noise realisation. In fact, one would expect their distribution in position to be uniform, as the noise is uncorrelated between noise realisations. Detections of the central cluster, however, should consistently be present in the reconstructions. Considering the number of times that the cluster is detected over an ensemble of noise realisations therefore gives a measure of the fidelity of the cluster detection, or the probability that we will be able to reconstruct that cluster from a given set of noisy shear data using the chosen noise threshold.

\begin{figure}
\includegraphics[width=0.45\textwidth]{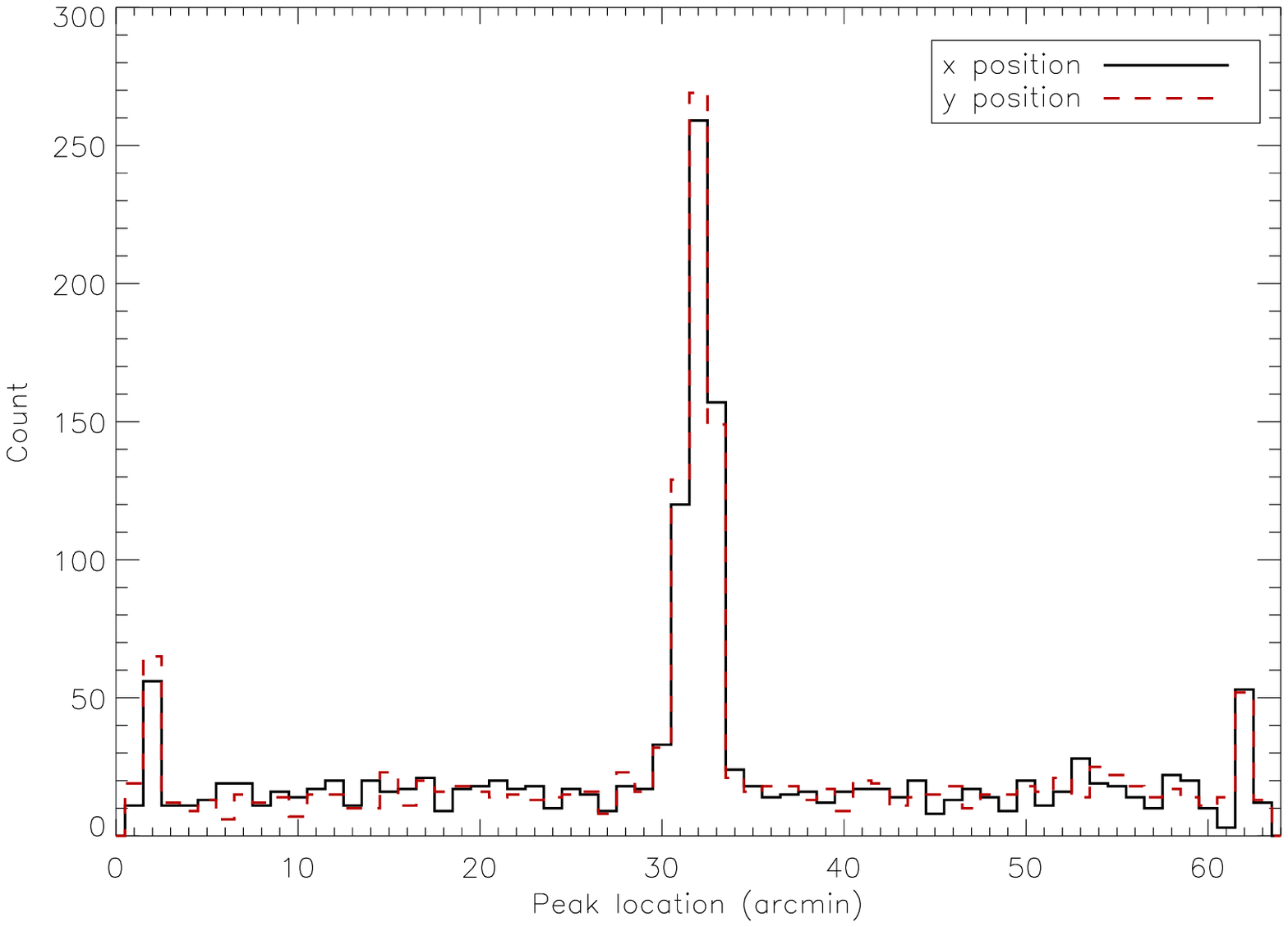}
\vskip4ex
\includegraphics[width=0.45\textwidth]{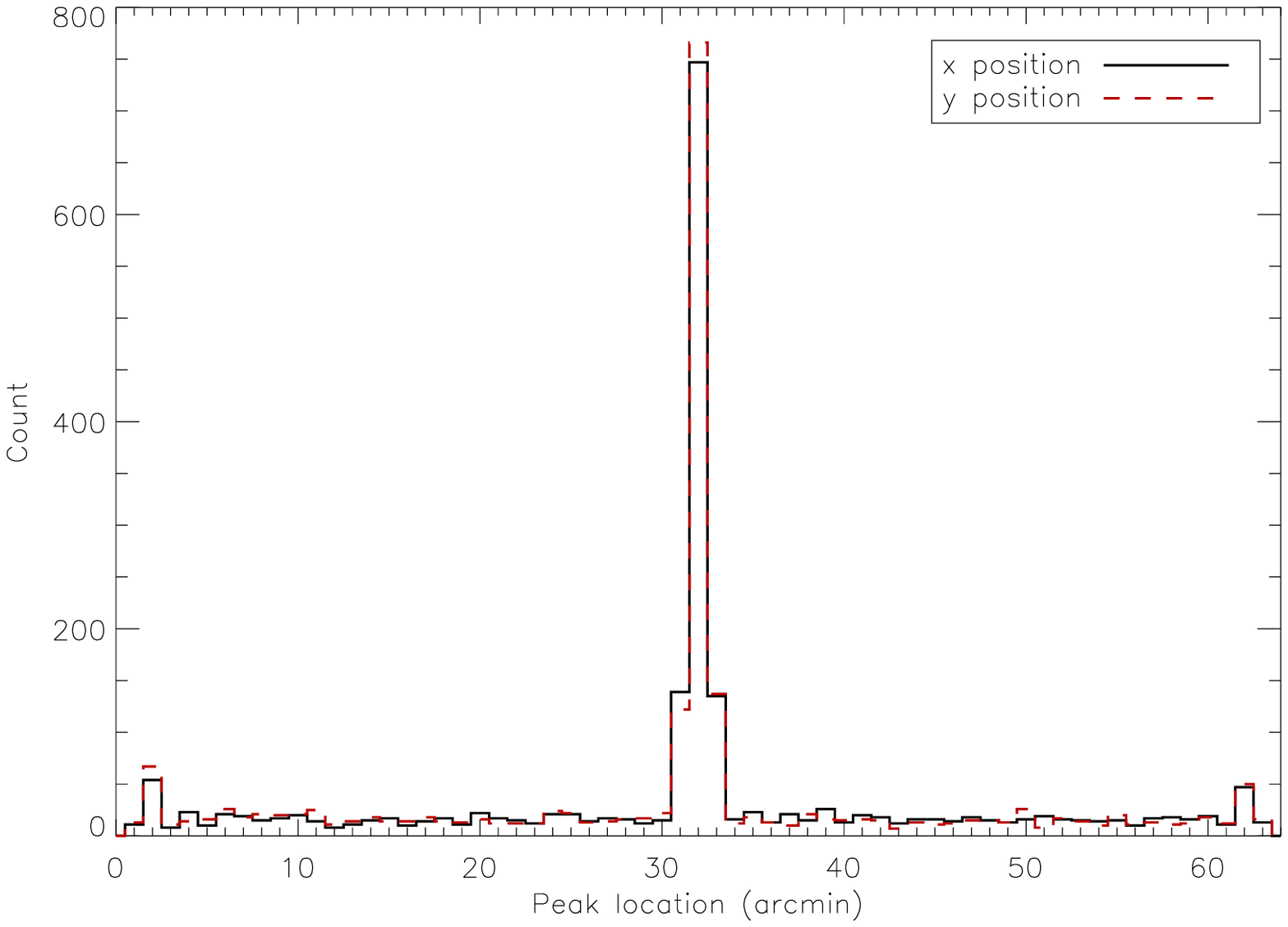}
\caption{The distribution of the $x-y$ locations of peaks detected in 1000 reconstructions of a given cluster field. Plotted are the distributions for field 48 (top panel, $M_{vir} = 2\times10^{14}h^{-1}M_\odot,\ z_{48}=0.35$) and field 53 (bottom panel, $M_{vir} = 7\times 10^{14}h^{-1}M_\odot,\ z_{53}=0.35$). In both fields, the central cluster is clearly identified, however the cluster in field 48 is less massive, and therefore the frequency of its detection is lower than that of the cluster in field 53. Away from the centre of the field, the distribution of (false) peaks is seen to be uniform, except near the edge of the field where a small overdensity is seen. \label{fg:xyz}}
\end{figure}

Figure \ref{fg:xyz} shows the distribution of detected peaks for two cluster fields. The top panel shows the distribution of $x$ and $y$ angular positions on the sky of all peaks detected in the 1000 realisations of the simulated data for field 48 ($M_{vir} = 2\times10^{14}h^{-1}M_\odot,\ z_{48}=0.35$), while the lower panel shows the distribution in $x$ and $y$ for the 1000 realisations of field 53 ($M_{vir} = 7\times 10^{14}h^{-1}M_\odot,\ z_{53}=0.35$).  The distribution of the peak positions is largely seen to be uniform over the field, but with a notable excess at $\{x,y\} = \{32^\prime,32^\prime\}$. There is also a slight excess seen near the edges of the field, which might be indicative of some systematic edge effects resulting from our method, possibly due to the computation of $P_{\gamma\kappa}$ and its inverse in Fourier space.

The $x$-$y$ peak distribution is well fit by a Gaussian plus a constant, with the width of the Gaussian consistently found to be in the range $0.5\lesssim [\sigma_x,\sigma_y] \lesssim 1.0$, except in two cases where no significant central detection was seen (fields 77 and 87, both with $M_{vir}=10^{14}h^{-1}M_\odot$; $z_{77}=0.65,\ z_{87}=0.75$). We therefore considered a detection of the cluster to be any peak found to lie within $30^\prime < [x,y] < 34^\prime$, with no constraint on the redshift at which the peak was detected. 

In rare cases, a high redshift detection is seen in the reconstruction in addition to a lower-redshift peak. This is similar to the high redshift false detections seen in LDS12, though it occurs far less frequently when using GLIMPSE. We opted to keep both peaks as cluster detections in all of the analysis which follows, as there is no truly blind way to discard one or other of the peaks as being a false peak. For this work, we wish to analyse the accuracy with which we can recover the mass and redshift of the simulated clusters, and to do this in an unbiased way, we cannot preferentially discard either detection based on our prior knowledge of the true location of the cluster. This will add to the noise in our redshift and mass estimates, but will ensure that we are not preferentially biasing our results, and thereby unfairly shrinking our error bars.

\begin{figure}
\includegraphics[width=0.45\textwidth]{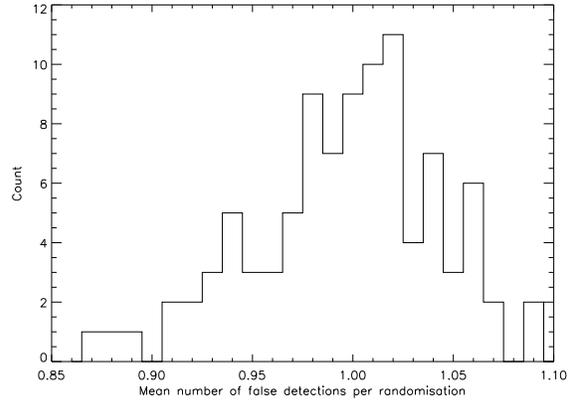}
\caption{The mean number of false detections per randomisation for each of the 96 simulated cluster fields. The histogram shows the number of cluster fields (the count) with a specified mean false detection rate. This mean is taken by considering the number of peaks detected in 1000 different reconstructions whose centroids lie outside of the central $4 \times 4$ pixels.\label{fg:ffalses}}
\end{figure}

Clearly, the number of false detections seen is indicative of the effectiveness of our algorithm at denoising the weak lensing measurements. A high threshold would be expected to remove all false detections within a reconstruction; however, extremely strict thresholding might remove part of the true lensing signal, meaning that the central cluster we aim to reconstruct might not be detected. Figure \ref{fg:ffalses} shows the mean number of false detections per reconstruction for each of the 96 simulated fields. This number is consistently around 1 per reconstruction, with no dependence seen on the mass and redshift of the simulated cluster in that field. Stricter thresholding would be expected to reduce this number; however, given that the position of these false detections is random, they can easily be identified as false detections by considering randomisations of the data, as they do not persist in reconstructions using different noise realisations.

\subsection{Cluster detection}

The number of times the central cluster is detected in the ensemble of noise realisations is indicative of both the fidelity of the detection of a given structure (i.e. whether that structure is a real structure or simply due to the noise) and diagnostic of the trade-off between noise removal and sensitivity to real structures. We expect that the probability of detecting a cluster halo will depend on both the mass of that halo and its redshift, for a fixed noise and threshold level. 

\begin{figure}
\includegraphics[width=0.45\textwidth]{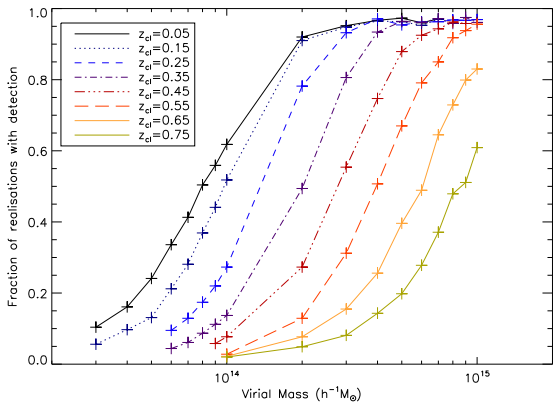}
\caption{Fraction of fields with a central cluster detection, $f_{det}$ as a function of cluster virial mass and redshift.\label{fg:detfracs}}
\end{figure}

Plotted in figure \ref{fg:detfracs} is the fraction of noise realisations (out of 1000) in which one or more central density peak was detected as a function of virial mass, with different coloured curves corresponding to cluster haloes at different redshifts. We denote this fraction in the analysis that follows by $f_{det}$. 

These curves effectively show the probability of reconstructing haloes of a given mass and redshift from a set of noisy data, given our choice of threshold parameter and shape noise level. Clearly, as clusters are moved to higher redshift, they become more difficult to detect; however at the high end of the mass function ($\gtrsim 8\times10^{14}h^{-1}M_\odot$), we can still expect to detect $\gtrsim60\%$ of clusters at redshift $z_{cl} = 0.75$, and this trend is expected to continue to higher redshifts, though with an increasing mass threshold. Lowering our denoising threshold would be expected to increase the completeness at all redshifts, allowing us to detect lower mass haloes, but will come at the cost of an increased number of false detections.

\subsection{Line-of-sight smearing}

One of the major issues identified in linear approaches to weak lensing reconstructions is a broad smearing of the structures along the line of sight. \cite{sth09} refer to this as a radial PSF, with a typical 1$\sigma$ width of $\sigma_{psf} \sim 0.2-0.3$. This very broad smearing makes it difficult to localise structures in redshift, and therefore difficult to study statistics of the density field as a function of redshift using the reconstructed maps.

LDS12 showed that a sparsity-based approach dramatically improves on this particular point; choosing a sparsifying dictionary that is comprised of Dirac delta functions along the line of sight should help to ensure that our reconstruction is sparse along the line of sight. GLIMPSE works under the same principle, whereby it seeks to find the most sparse reconstruction that is compatible with the data. 

\begin{figure}
\includegraphics[width=0.45\textwidth]{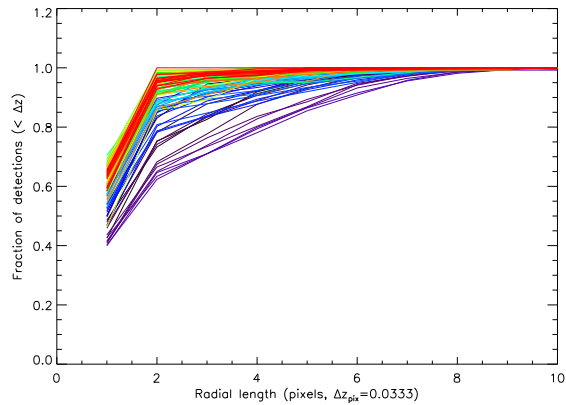}
\caption{Fraction of detections for a given cluster field that are found to have a radial extent $\le \Delta z$. The different coloured lines represent different cluster fields. Given our pixel size ($\Delta z_{pix} = 0.033$), this smearing is substantially smaller than that seen in the method of Simon et al. (2009), who found a typical smearing with standard deviation $\sigma \sim 0.2-0.3$. \label{fg:smearing}}
\end{figure}

Figure \ref{fg:smearing} shows the distribution of the total radial length of central structures detected in our reconstructed fields. The length plotted is not a 1$\sigma$ width, but rather the full radial extent of the structure where $\delta>1$. For each cluster field, we plot the fraction of detections with a width $< \Delta z$ as a function of $\Delta z$. In all fields, more than 50\% of the detections have a width less than or equal to $\Delta z = 2\,\rm{pix} = 0.067$. Line-of-sight smearing of the order $\Delta z \sim 0.2$ ($\sim 6\,{\rm pix}$) is clearly very rare in our reconstructions, and the largest smearing seen ($10\,{\rm pix}, \Delta z \sim 0.33$) is still smaller than the typical spread seen in the \citeauthor{sth09} method.

\subsection{Redshift estimation}
\label{subsec:zest}

We now consider the accuracy of our reconstructions, by first assessing how closely the redshift of a reconstructed cluster matches that of the original simulation. To do this, we consider the distribution in redshift of any peaks whose centroid lies within the central $4\times 4$ angular pixels of the image in the 1000 reconstructions of that cluster field. This redshift was computed by first identifying the pixels associated with the reconstructed peak that had a density contrast $\delta \ge 1$. We then computed the $\delta$-weighted centroid in $x$, $y$ and $z$. 

For each cluster field, we then considered the distribution in the estimated redshift over the ensemble of noise realisations. We consider two redshift estimators for the cluster. The first is the median redshift in this distribution, for which error bars were computed by finding the redshift range, centred on the median value $z_{med}$, that encompassed $68.2\%$ of the redshift estimates (thereby representing the $1\sigma$ confidence interval of the redshift estimate). 

We also estimated the probability distribution function (PDF) for the redshift estimates using an adaptive kernel density estimate (AKDE) with a standard Epanechnikov kernel \citep{silverman86}. From this, we estimated cluster redshift $z_{peak}$ as the peak in the PDF. A $1\sigma$ confidence interval was then computed from the cumulative distribution function obtained by integrating the PDF, always ensuring that 68.2\% of the estimated redshift values fell within this interval. 

\begin{figure*}
\includegraphics[width=0.25\textwidth]{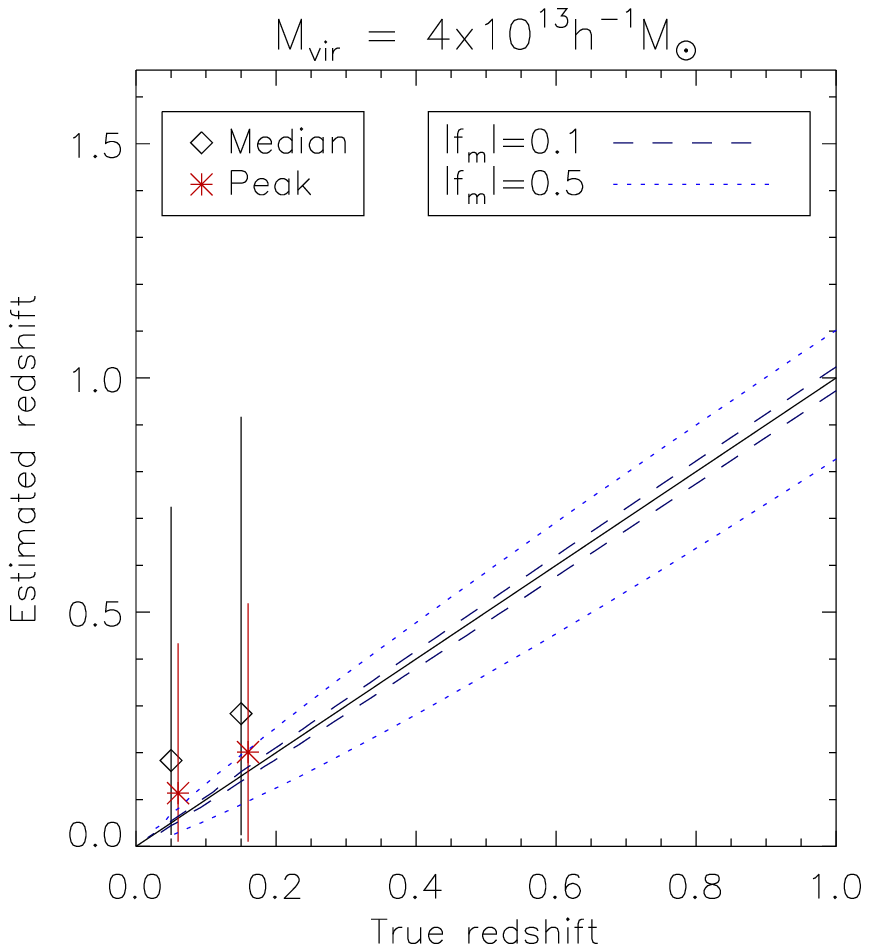}\includegraphics[width=0.25\textwidth]{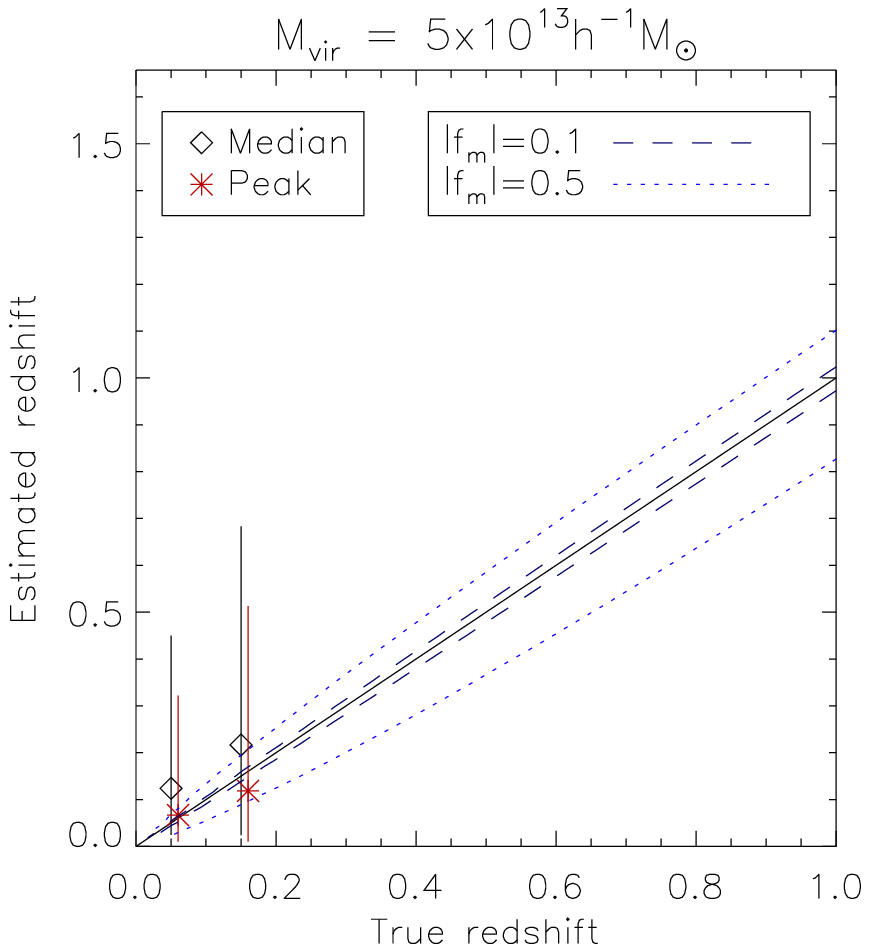}\includegraphics[width=0.25\textwidth]{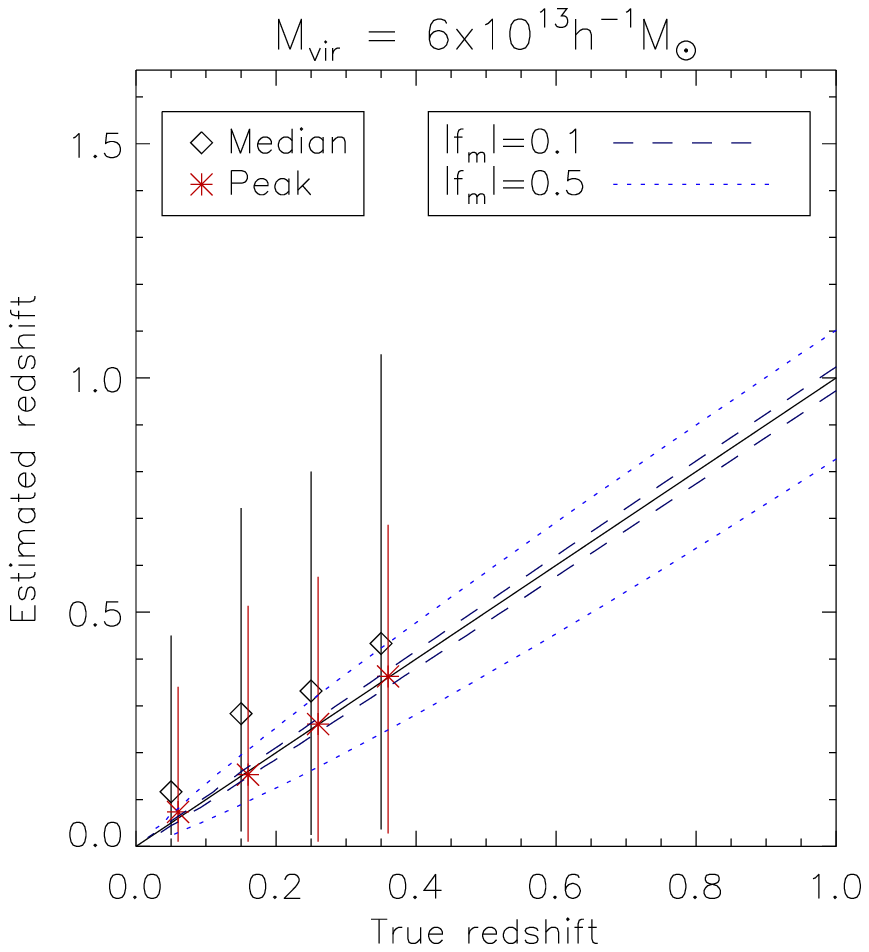}\includegraphics[width=0.25\textwidth]{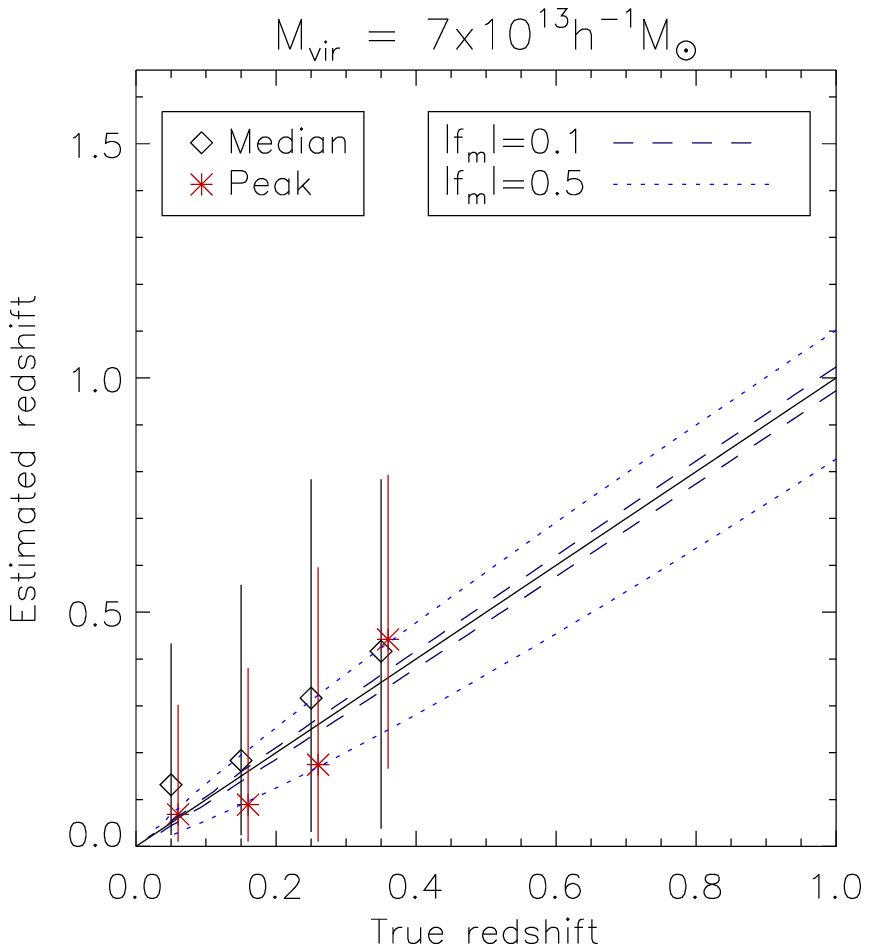}
\vskip2ex
\includegraphics[width=0.25\textwidth]{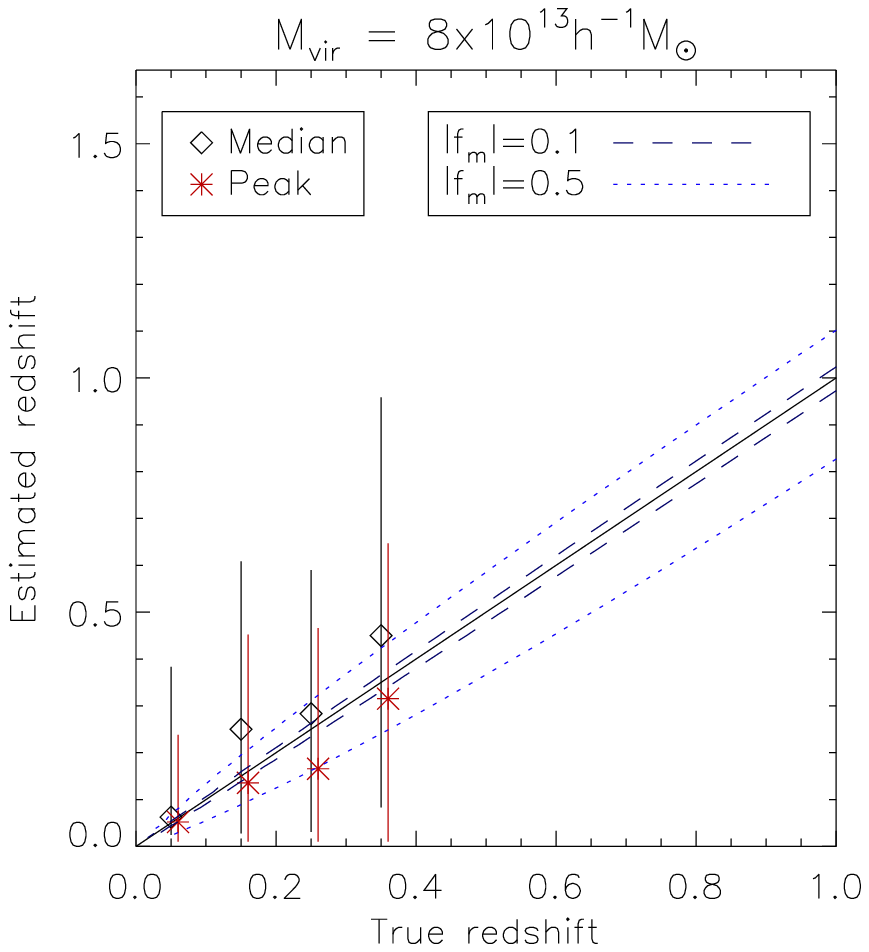}\includegraphics[width=0.25\textwidth]{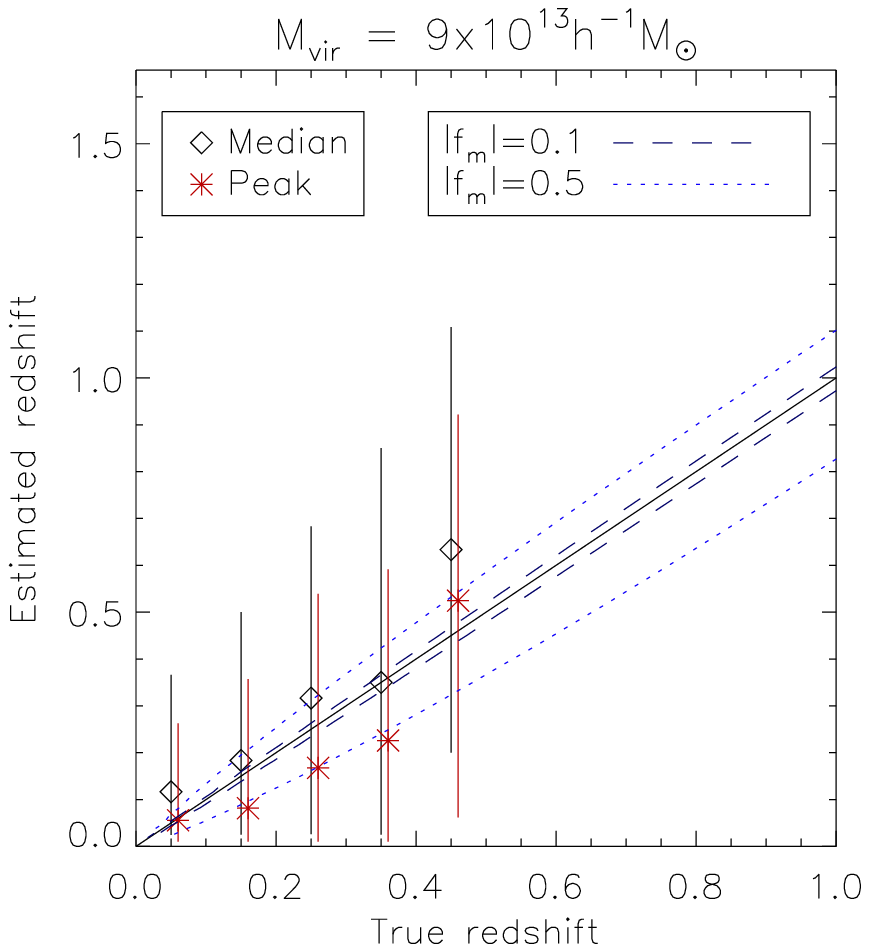}\includegraphics[width=0.25\textwidth]{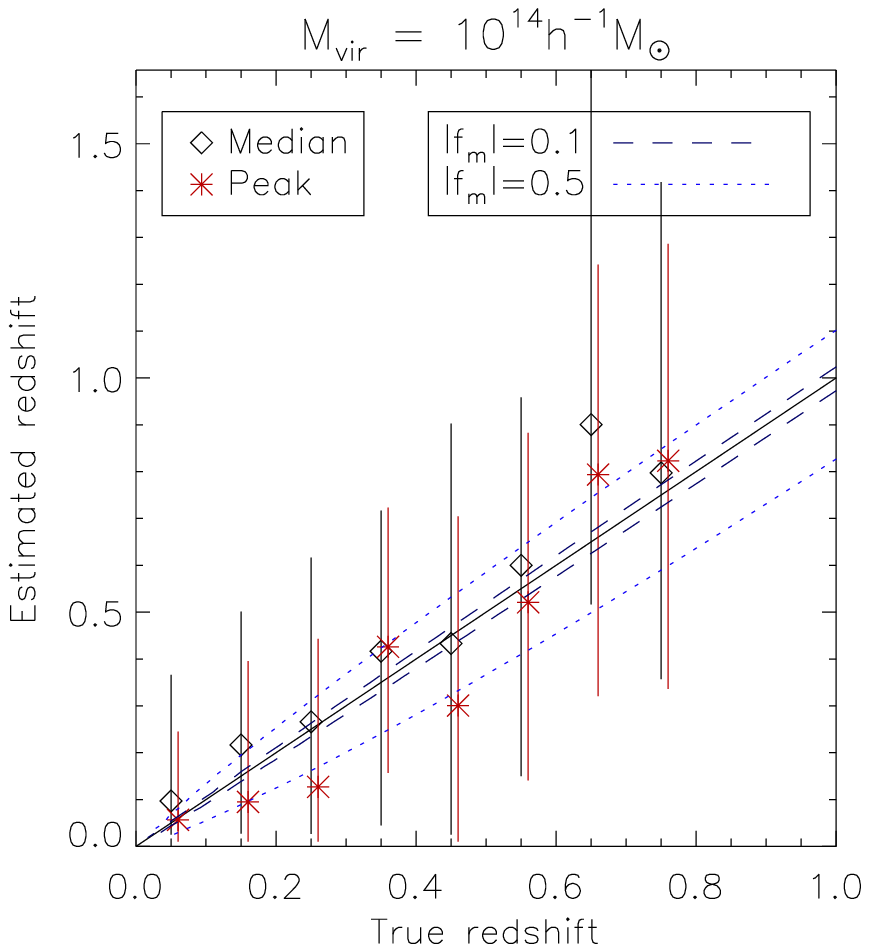}\includegraphics[width=0.25\textwidth]{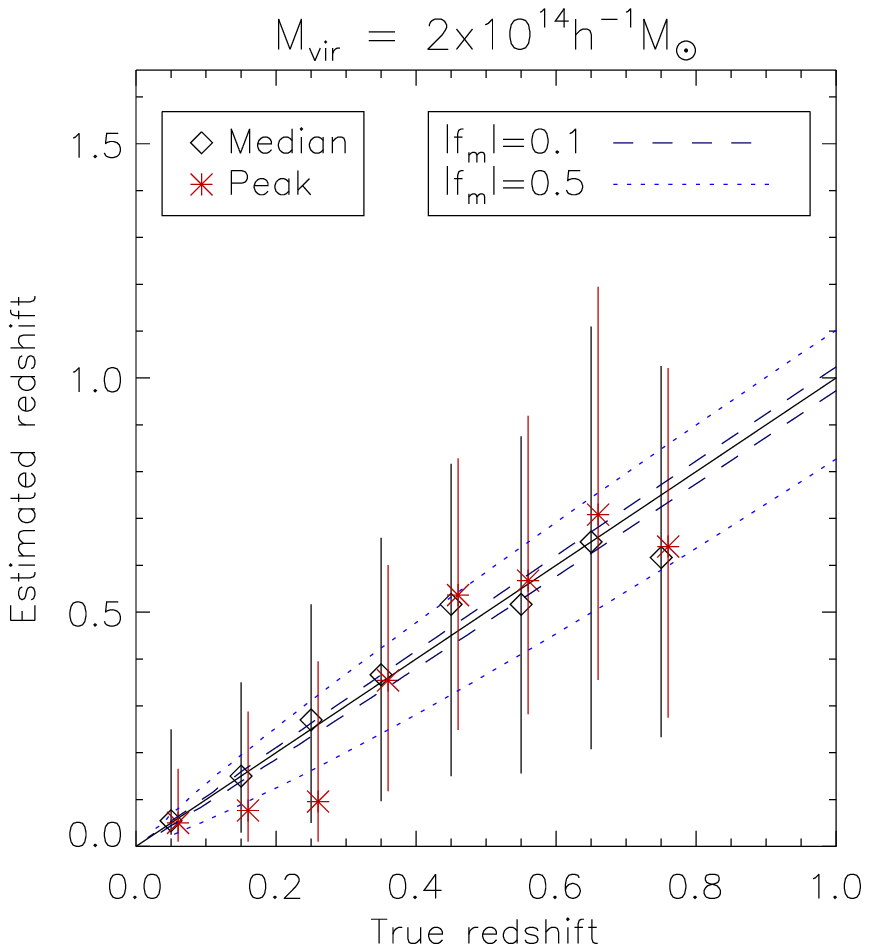}
\vskip2ex
\includegraphics[width=0.25\textwidth]{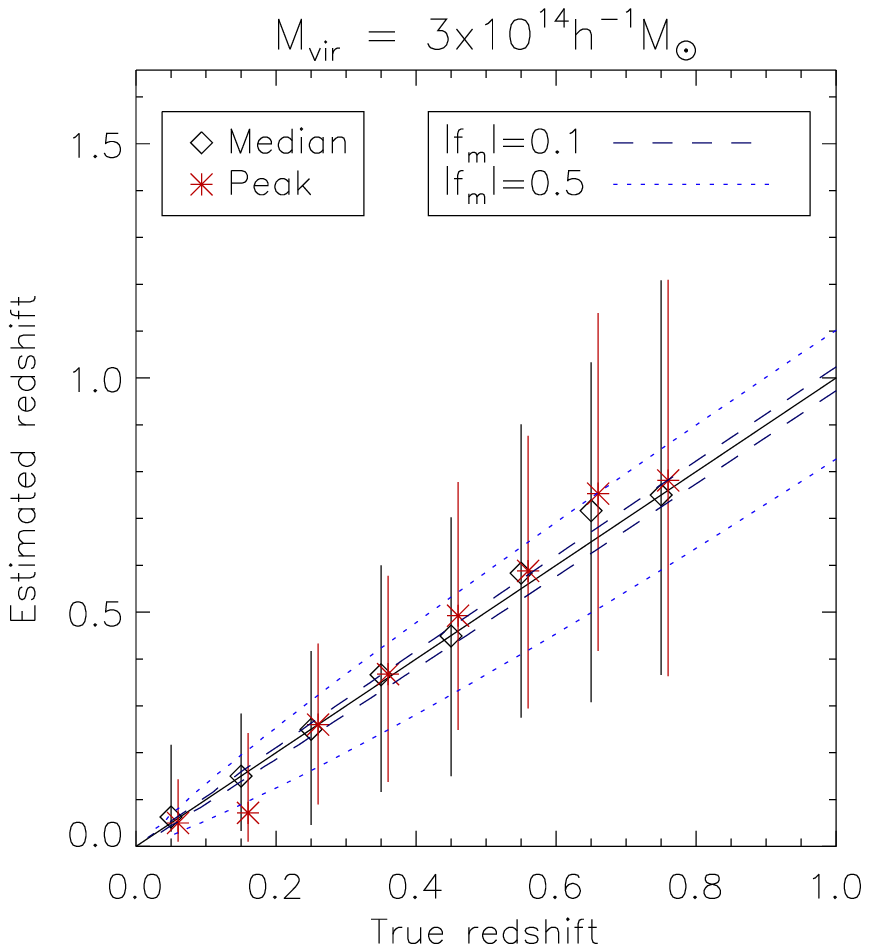}\includegraphics[width=0.25\textwidth]{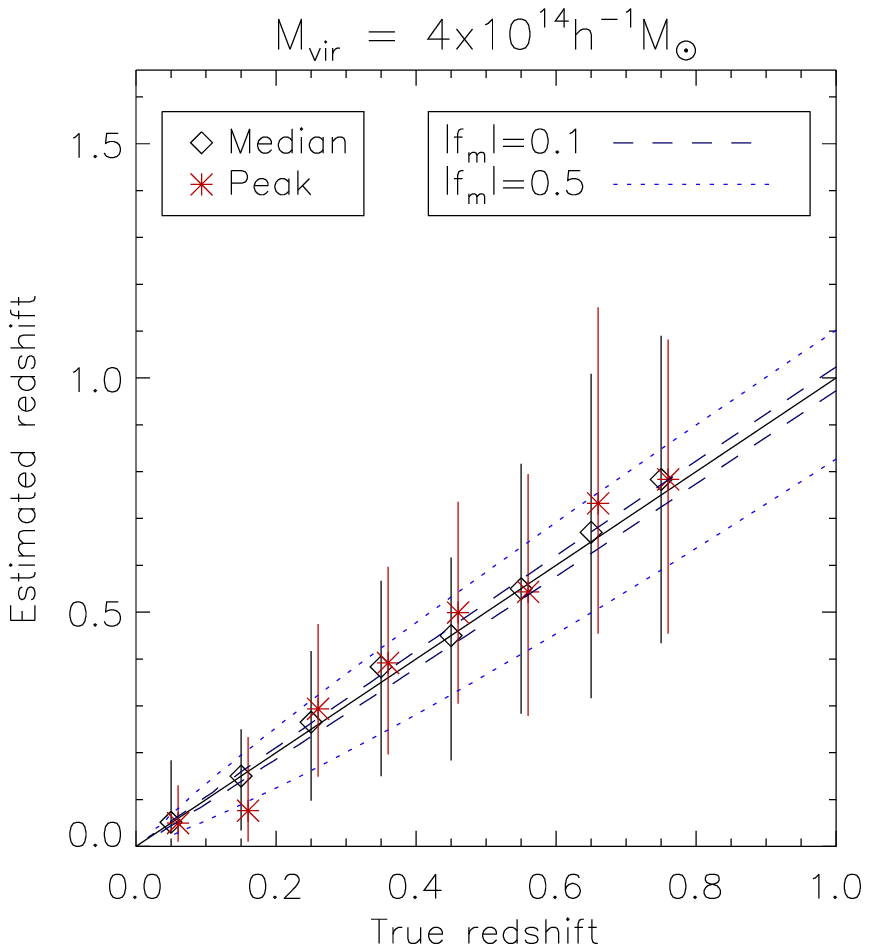}\includegraphics[width=0.25\textwidth]{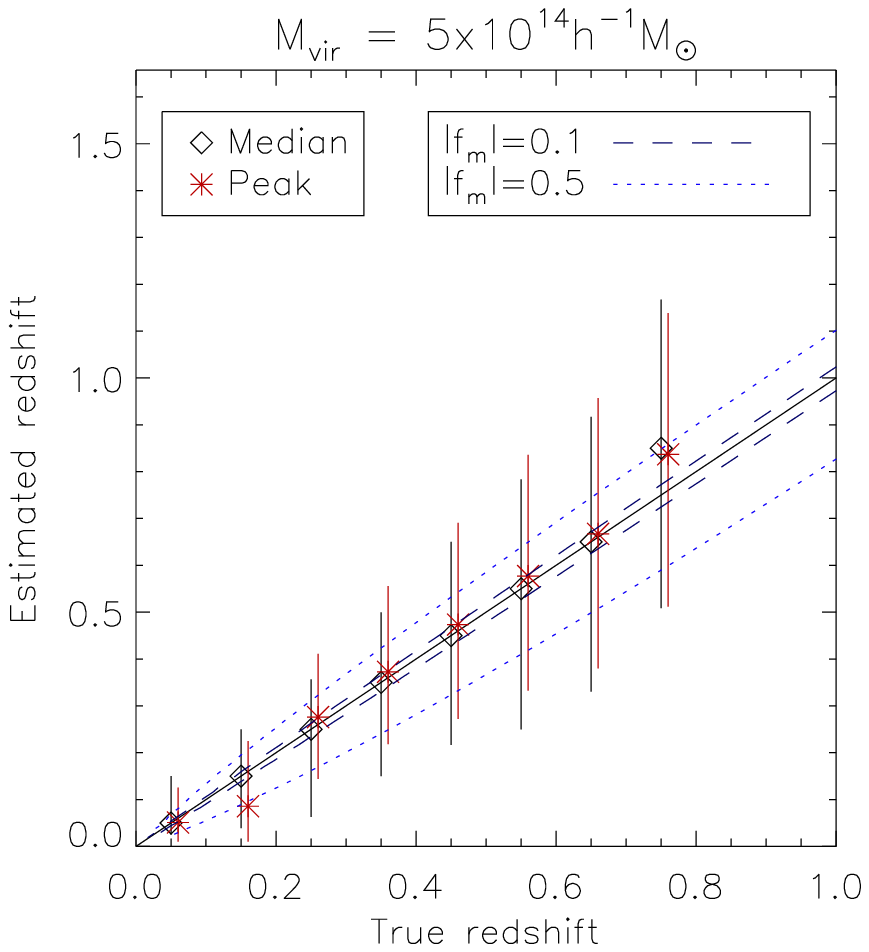}\includegraphics[width=0.25\textwidth]{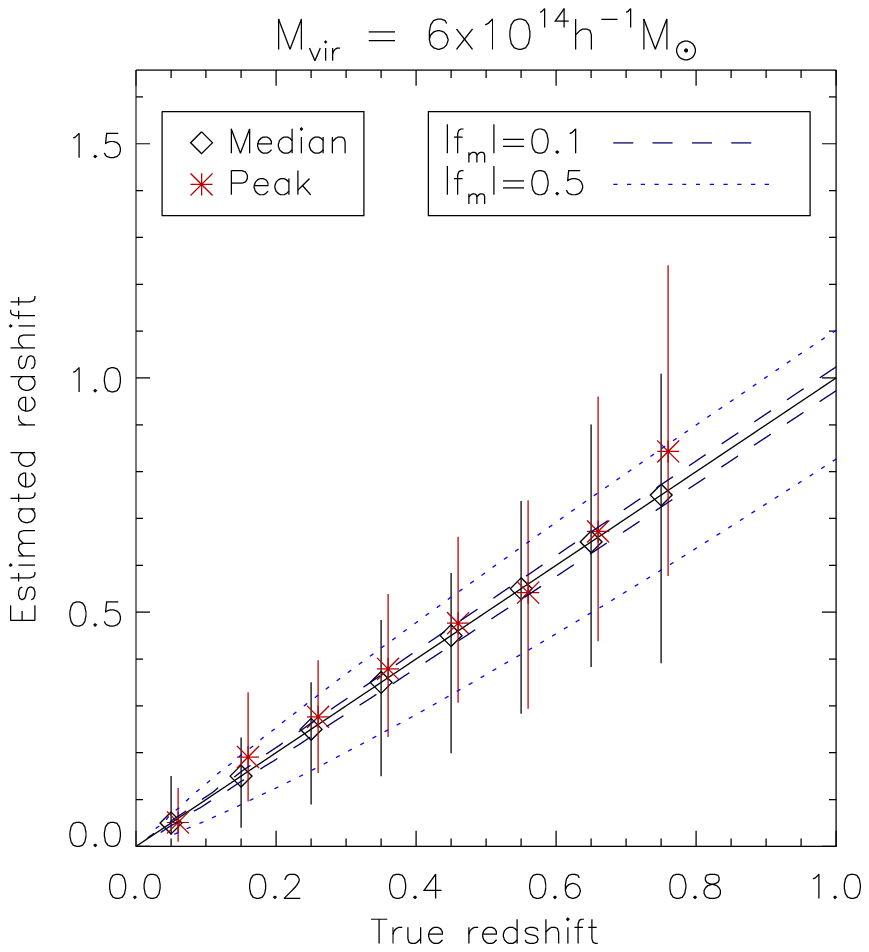}
\vskip2ex
\includegraphics[width=0.25\textwidth]{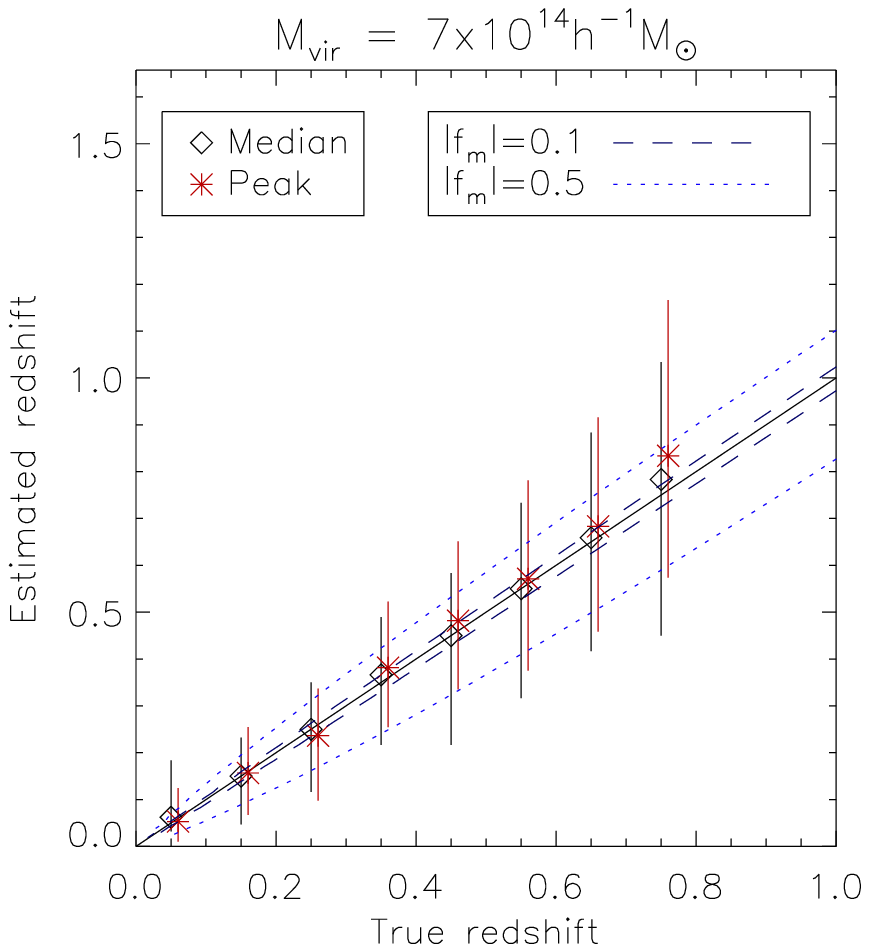}\includegraphics[width=0.25\textwidth]{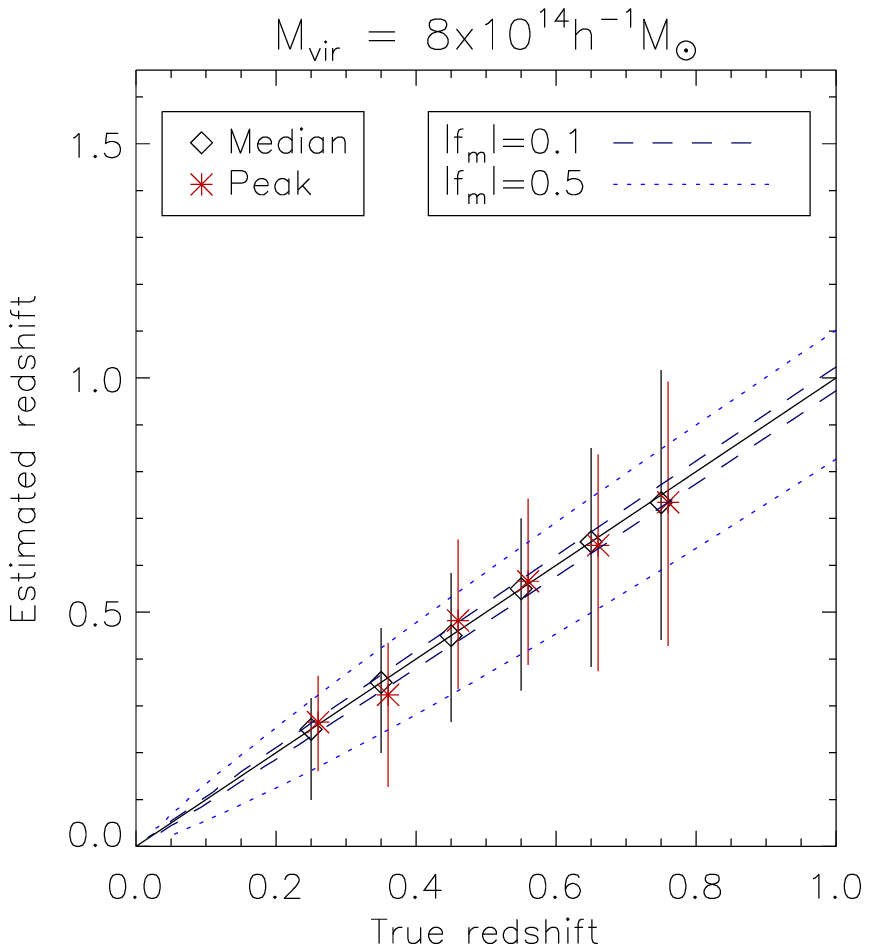}\includegraphics[width=0.25\textwidth]{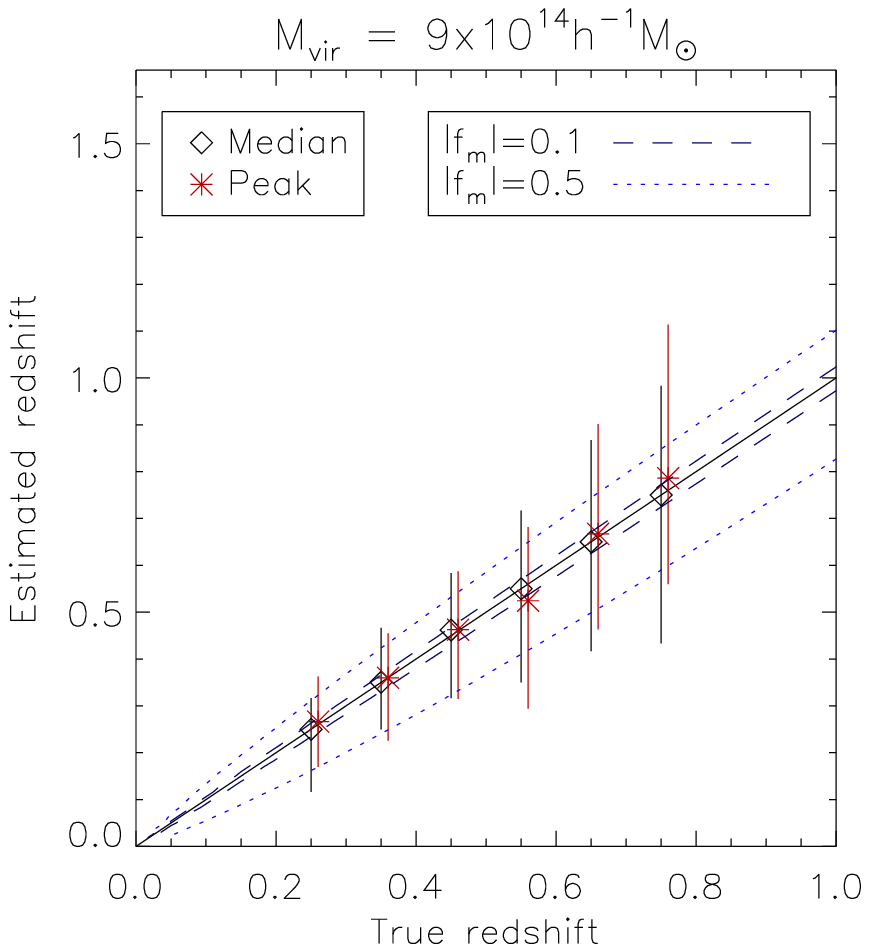}\includegraphics[width=0.25\textwidth]{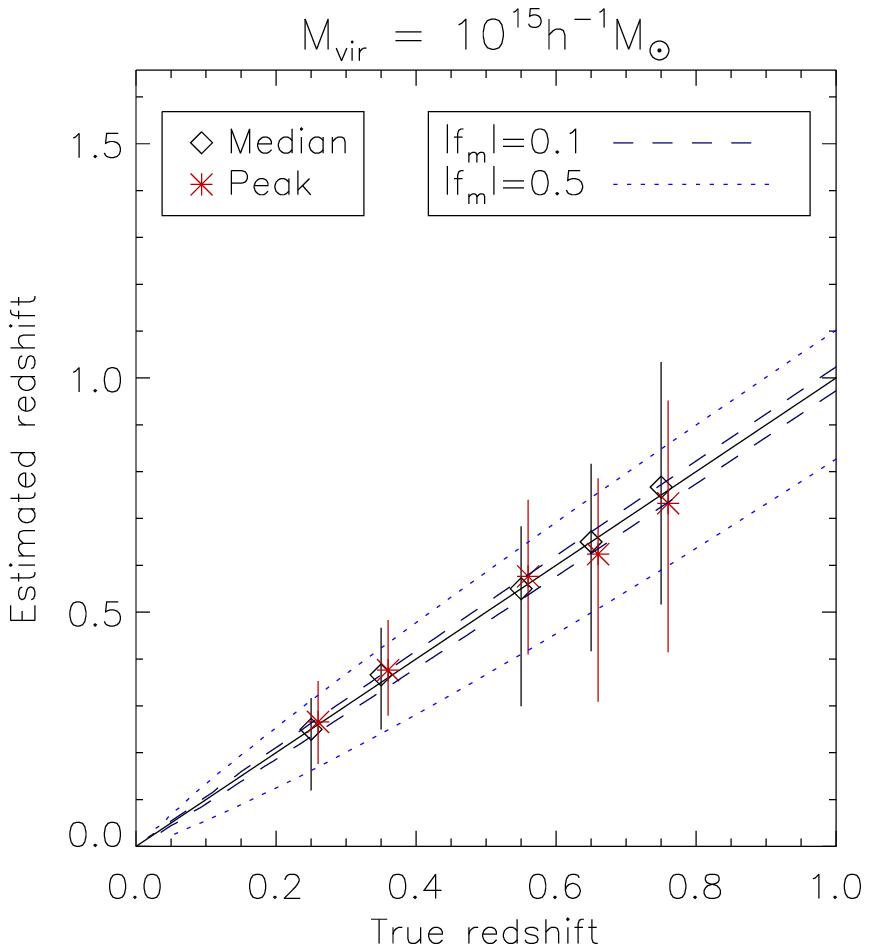}
\caption{Estimated vs true redshift for the 96 clusters in our sample, separated by cluster virial mass. Red points indicate the peak of the AKDE distribution, while black points indicate the median value considering all noise realisations. The red points have been shifted by 0.01 in both directions for visualisation purposes. \label{fg:zests}}
\end{figure*}

Figure \ref{fg:zests} shows the estimated vs true redshift for clusters in our sample as a function of mass. Note that the results for $M_{vir}=3\times10^{13}h^{-1}M_\odot$ are not plotted in the interests of space, but are consistent with the $M_{vir}=4\times 10^{13}h^{-1}M_\odot$ clusters. Shown in the figure are the median estimate and the AKDE peak redshift, the latter being offset by $\delta z = 0.01$ in both directions for visualisation purposes. Both estimates clearly yield an accurate estimate of the redshift, with the peak estimate seen to have smaller error bars in many cases. No systematic bias is seen in the peak case, while the median estimate does appear to slightly overestimate the redshifts for low-mass haloes. The error bars are seen to increase in size with increasing redshift, and with decreasing halo mass, as expected.

As will be discussed more fully in section \ref{subsec:massest}, an error in redshift will naturally give rise to a bias in the estimated mass of the cluster. We can parameterise this error through the relation $M(z_{est}) = (1\pm |f_m|)M(z_{true})$. The contours labelled $|f_m| = 0.1$ and $|f_m| = 0.5$ refer to the redshift ranges within which the fractional error on the mass is less than $10\%$ and $50\%$, respectively. The majority of the median redshift estimates fall within the contours of $|f_m| = 0.1$. 

\subsection{Mass estimation}
\label{subsec:massest}

Given our ability to accurately estimate the redshifts of clusters using GLIMPSE, we now consider how accurately we might be able to estimate the masses of the haloes detected. Estimating the masses of clusters detected using our method is complicated by several factors, however.

Firstly, for any given noise realisation, the cluster will be detected at a given redshift $z_i$. The density contrast is related to the mass by
\begin{equation}
M_n = \sum_n \delta^{(n)}(z_i)\overline{\rho}(z_i)V_{pix}(z_i)\ ,
\end{equation}
where $\overline{\rho}(z_i)$ is the mean matter density at redshift $z_i$ and $V_{pix}(z_i)$ is the comoving volume covered by one pixel in our reconstruction at that redshift, and the mass $M_n$ is the mass enclosed by $n$ angular pixels. Therefore, the estimated mass depends on redshift through both $\overline{\rho}$ and $V_{pix}$, and any redshift error in a given reconstruction will bias the mass estimate by a corresponding factor. If the cluster is detected at a different redshift, then for the same amplitude in $\delta$, the estimated mass will change proportionately as $\overline{\rho}V_{pix}$.

Moreover, the density contrast estimate itself will be biased if the cluster is placed at the wrong redshift. Consider the relationship between the measured shear and the density contrast in equation \eqref{eq:gamdelta}: $\gamma = \mathbf{P}_{\gamma\kappa}\mathbf{Q}\delta$. The operator $P_{\gamma\kappa}$ has unit norm, so the normalisation in $\gamma$ and $\kappa$ is the same. However, this is not the case for $\mathbf{Q}$, where the rows and columns of the matrix are functions of the redshift.

Recall that $\mathbf{Q} \propto D_{ls}/D_{s}$. This means that if a cluster is detected, for example, at a higher redshift, it will be reconstructed with a larger amplitude in order to compensate for the decrease in the estimated $D_{ls}/D_{s}$. This scaling is encoded by the normalisation matrix $\mathcal{N}$ defined in equation \eqref{eq:normalisation}, ${\mathcal N}(z_n)\equiv{\mathcal N}_{nn} = \sum_p \mathbf{Q}_{pn}^2$, where the index $n$ refers to a given redshift slice. As the normalised density contrast, the quantity we reconstruct with GLIMPSE, is related to $\delta$ via $\delta^\prime = \mathcal{N}\delta$, the density contrast estimated at a given redshift (and thereby the estimated mass at that redshift) is scaled by $M_{n}(z) \propto \delta(z) \propto {\mathcal N}^{-1}(z)$.

The estimated mass at redshift $z_{est}$ is therefore related to the mass estimate at the true redshift $z_{true}$ via:
\begin{equation}
\frac{M_n(z_{est})}{M_n(z_{true})} = \frac{\overline{\rho}(z_{est})}{\overline{\rho}(z_{true})} \frac{V_{pix}(z_{est})}{V_{pix}(z_{true})}\frac{{\mathcal N}(z_{true})}{{\mathcal N}(z_{est})}\ .
\label{eq:bias}
\end{equation}

\begin{figure}
\includegraphics[width=0.45\textwidth]{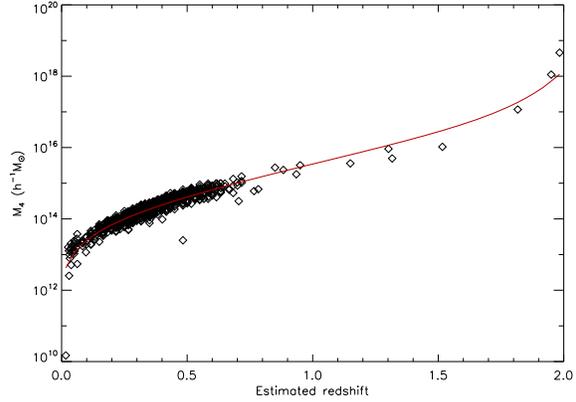}
\caption{The mass estimate within the central four angular pixels $M_4$ as a function of estimated redshift for the reconstructions of cluster field 53. The red curve shows $M_4$ measured from the input simulation, and rescaled with redshift according to the relation in equation \eqref{eq:bias}.\label{fg:mzdep}}
\end{figure}

In figure \ref{fg:mzdep}, we plot the estimated mass over the central four pixels $M_4$ for all the reconstructions of field 53 ($M_{vir} = 7\times 10^{14}h^{-2}M_\odot,\ z_{53}=0.35$), computed simply by multiplying the integrated density contrast by the mean density and the pixel volume at the redshift of detection, without accounting for any biases. It is very clear that there is a strong dependence of the mass estimate on the redshift. The red curve in the figure shows $M_4$ estimated on the true density contrast map of the cluster, scaled as a function of redshift according to equation \eqref{eq:bias}. 

This curve shows an excellent fit to the measured data, and we can therefore use the relationship in equation \eqref{eq:bias} to rescale the mass estimates for a given cluster onto the same redshift, to allow us to compare their values, and to compute an overall mass estimate for the cluster, and error bars associated with this estimate.

In order to assess the ability of our method to constrain the angular profile of a cluster, we consider the mass within the central $2\times 2$ pixels ($M_4$), $4\times4$ pixels ($M_{16}$), $6\times6$ pixels ($M_{36}$) and $8\times 8$ pixels ($M_{64}$). For each detected cluster, the density contrast was rescaled by ${\mathcal{N}}$, and then summed within the central $n$ angular pixels at all redshifts associated with the cluster. We then compute the mass assuming the true redshift of the cluster. As none of these mass measures directly corresponds to the virial mass of the simulated cluster, we also compute these values directly from the simulated maps for comparison.

\begin{figure*}
\includegraphics[width=0.25\textwidth]{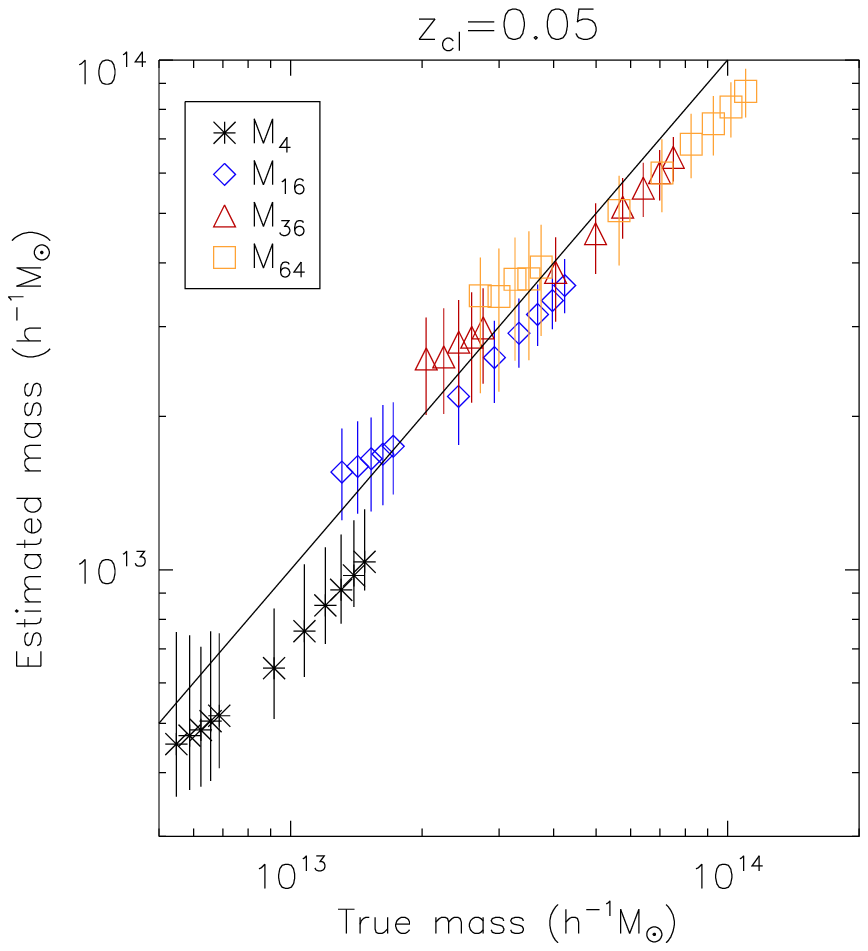}\includegraphics[width=0.25\textwidth]{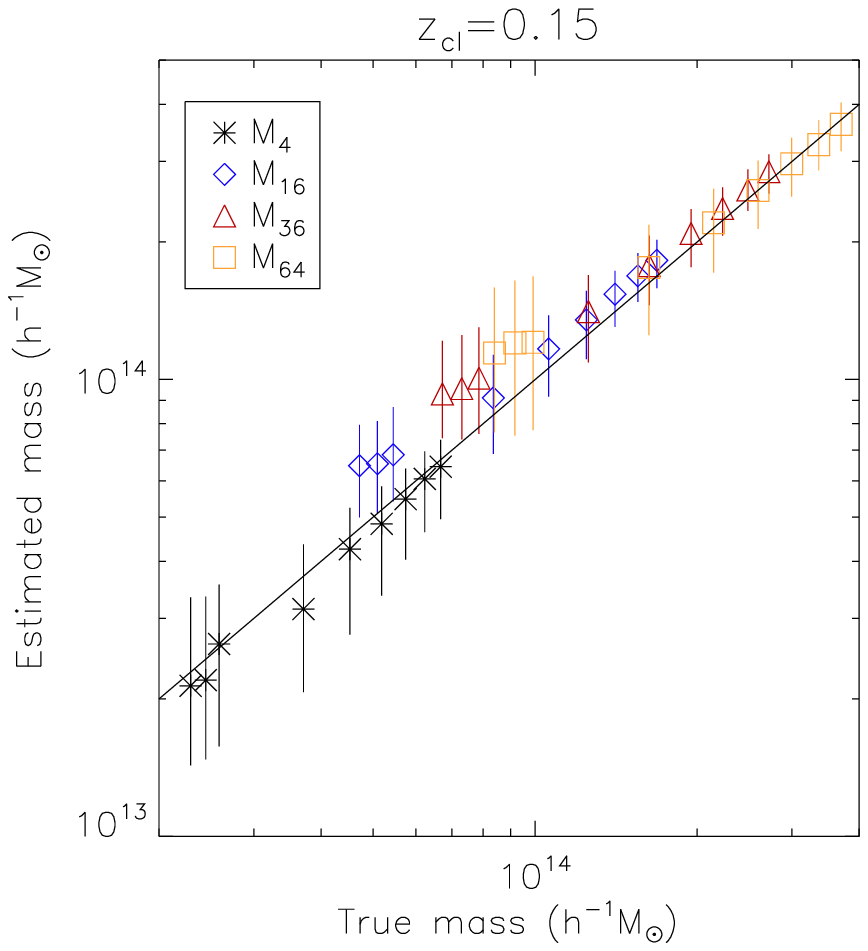}\includegraphics[width=0.25\textwidth]{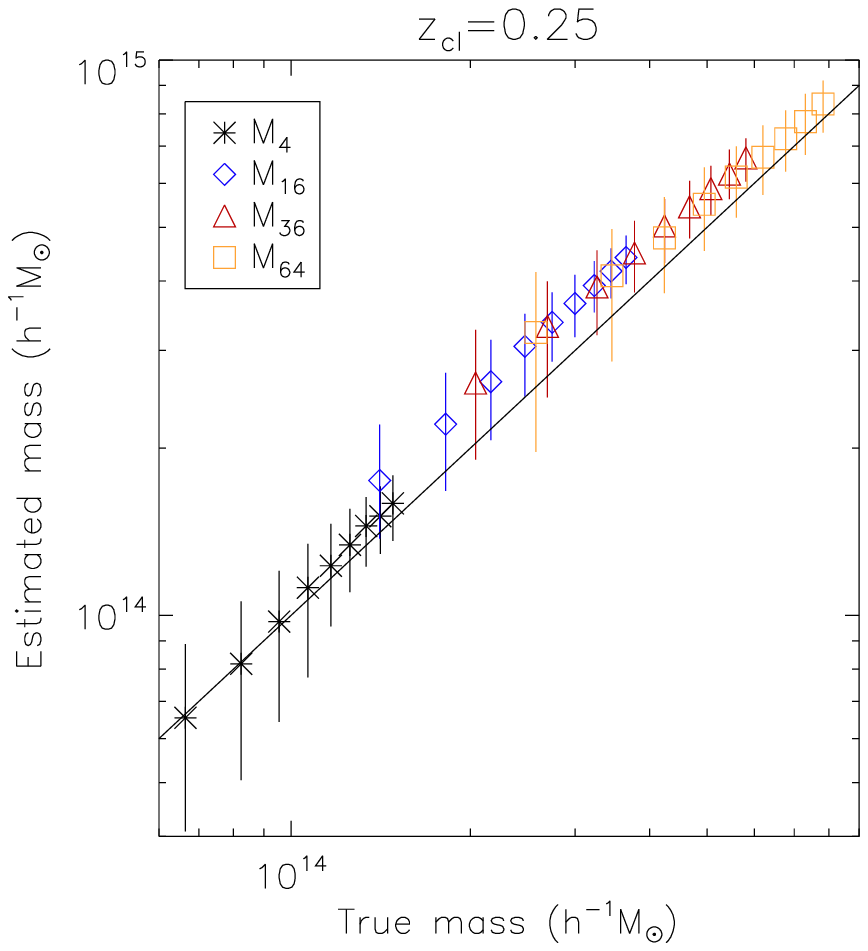}\includegraphics[width=0.25\textwidth]{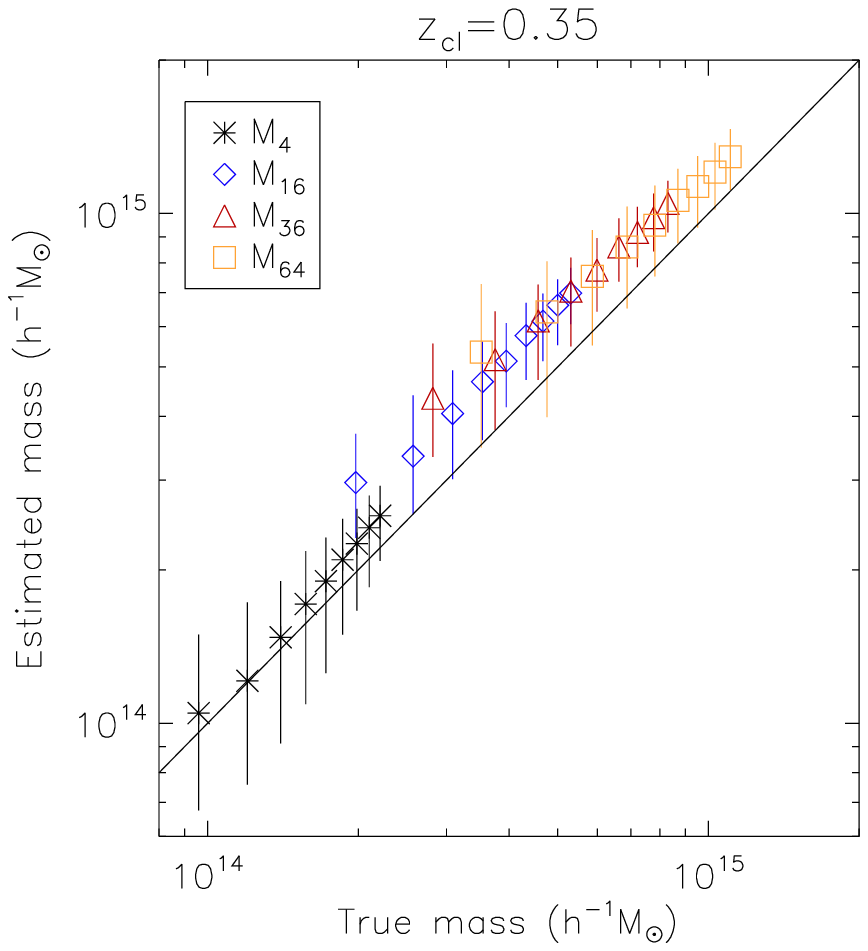}
\vskip4ex
\includegraphics[width=0.25\textwidth]{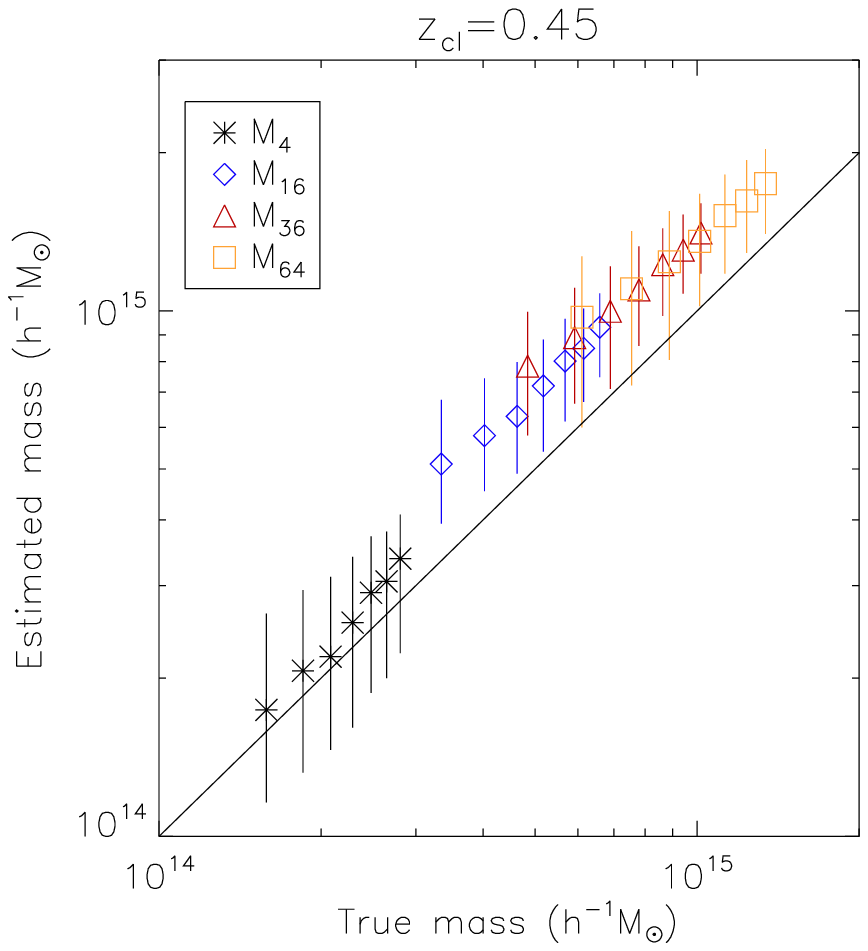}\includegraphics[width=0.25\textwidth]{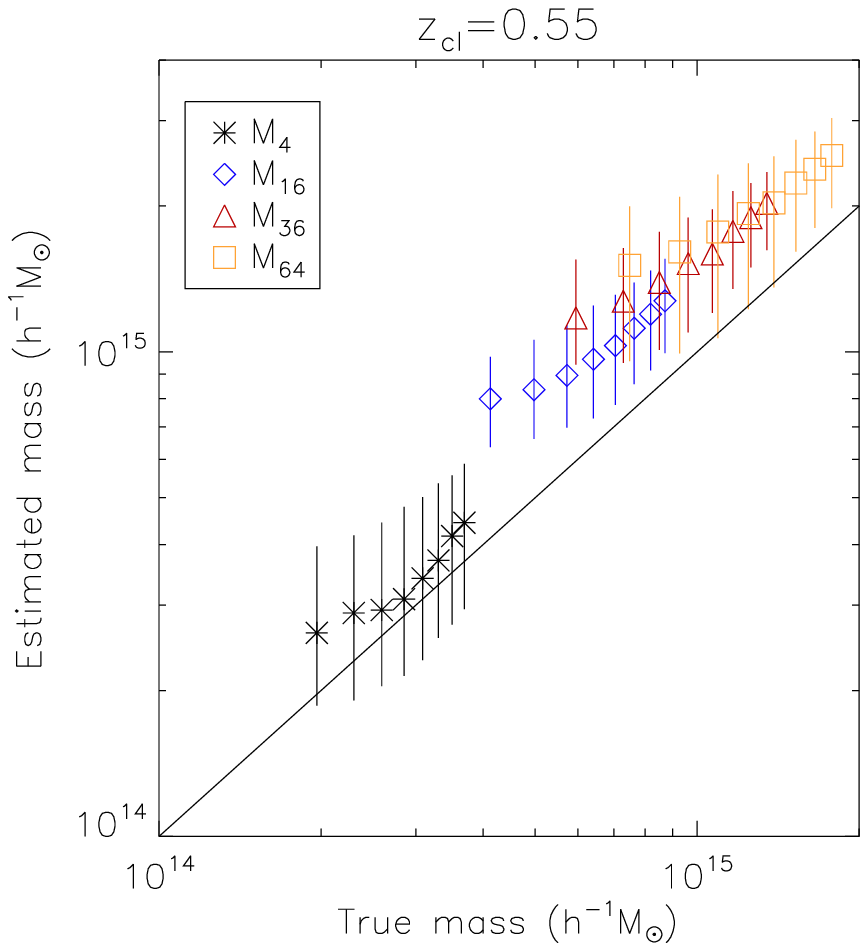}\includegraphics[width=0.25\textwidth]{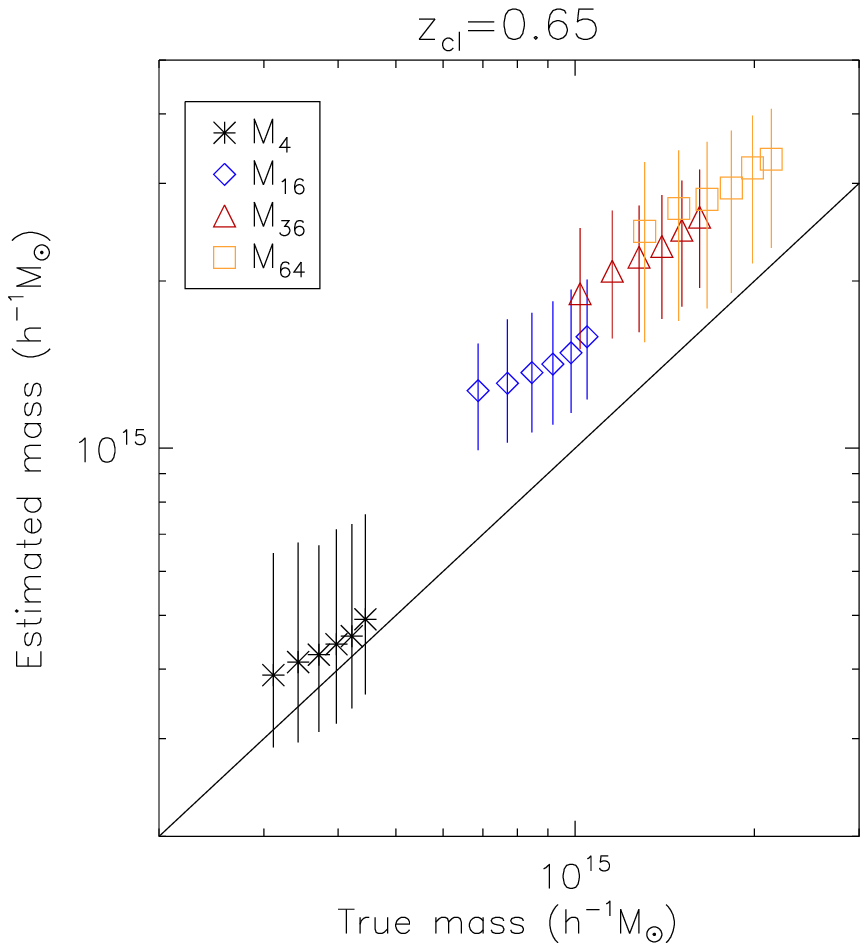}\includegraphics[width=0.25\textwidth]{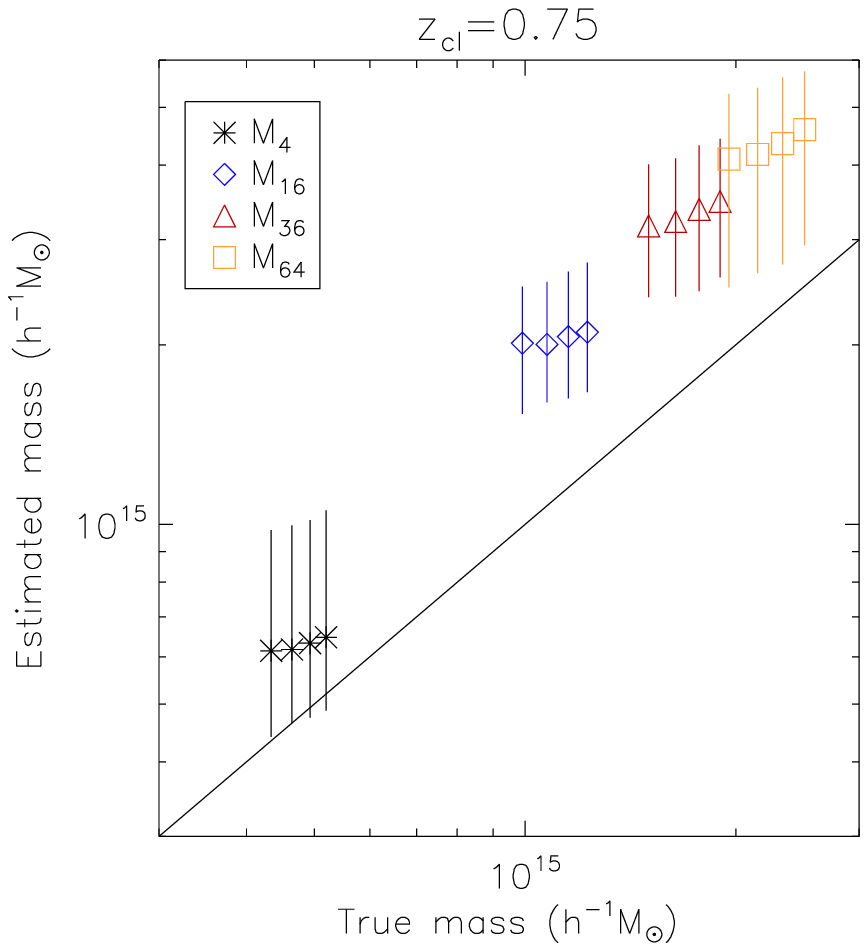}
\caption{Mass estimates at four different integration scales plotted against the true mass measured from the simulation. The mass estimates use the median rescaled mass over all the realisations of the field, and the error bars are $68.2\%$ confidence intervals computed symmetrically about the median value. \label{fg:ms}}
\end{figure*}

Given the distribution of estimated masses across all the noise realisations, we can then compute the median and $68.2\%$ confidence regions as for the redshift. Figure \ref{fg:ms} shows our results. Included are only results for clusters in which the fraction of detections $f_{det}>0.3$ (i.e. for which we have more than 300 detections out of the 1000 randomisations). This limit was placed to ensure that we have adequate sample points for our derived statistics. In figure \ref{fg:ms}, the four different estimators are shown on the same plot, and the clusters are divided up by redshift. 

At the lowest simulated cluster redshift of $z_{cl} = 0.05$, we see a tendency to underestimate the masses at all scales, while the opposite trend is seen at high redshift. However, in most cases, and particularly for $M_4$, this is at most a $1\sigma$ effect. In the intermediate redshift regime $0.15\le z_{cl} \le 0.55$, excellent agreement is seen in $M_4$. The trend to overestimate the mass at larger radii implies that our fit to the angular profile of the cluster is not perfect; however, the deviations seen remain at the $1\sigma$ level. 

We can conclude from these results that we are able to accurately recover the masses of most clusters in our sample, even at high redshift, with very little bias seen. However, this analysis has relied on the assumption that the true redshift of the cluster is known, which it will not be in real data. 

\begin{figure}
\includegraphics[width=0.45\textwidth]{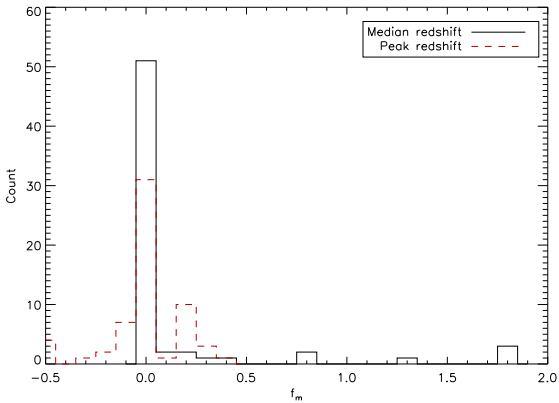}
\caption{Mass bias associated with a redshift error for the clusters included in figure \ref{fg:ms}, shown for both the median redshift estimate and the peak AKDE estimate. The bias is generally very low for the median estimate, with $|f_m|\le 0.05$ in most cases. The peak redshift estimate shows a wider distribution, but no extreme outliers with $|f_m| > 0.5$.\label{fg:bias}}
\end{figure}

Using equation \eqref{eq:bias}, we can compute the expected bias on the mass that will be obtained by assuming the median or peak AKDE estimate of the redshift, instead of the true redshift. As in section \ref{subsec:zest}, we model this as
\begin{equation}
\frac{M(z_{est})}{M(z_{true})} = 1 + f_m\ .
\end{equation}
equation \eqref{eq:bias} then directly gives us a measure of $f_m$. In figure \ref{fg:bias} we show the histogram of $f_m$ for the clusters included in figure \ref{fg:ms} (i.e. all the clusters with $f_{det}>0.3$). We plot this mass bias factor for the case in which we assume as the cluster redshift the median estimate (solid black line) and the peak AKDE redshift (dashed red line). In the case of the median estimator, the vast majority of clusters fall within the $\pm 0.05$ region, implying a smaller than $5\%$ induced bias from errors in the redshift estimate. The peak AKDE redshift shows a wider spread in $f_m$ values, but in contrast to the median redshift does not show any extreme outliers of $|f_m|>0.5$.

\section{Summary and discussion}
\label{sec:summary}

In this paper, we have presented GLIMPSE, a new sparsity-based approach to weak lensing density mapping in 3D. We have carried out 96 sets of cluster simulations, each containing a single cluster of a given mass and redshift, with 1000 different realisations of Gaussian shape noise at a level expected from current and upcoming weak lensing surveys. We included Gaussian photometric redshift errors in these simulations, again at a level typically expected with current experiments. In all our reconstructions, we improved the resolution in the redshift direction by a factor of $6$ compared with the input data, and our resulting reconstructions had a redshift bin width of $\Delta z_{pix} = 0.033$.

We demonstrated an excellent precision in the localisation of the detected cluster in the $x-y$ plane, and showed that the incidence of false detections in our reconstructions is $\sim 1$ per reconstruction, with the location of this false peak being entirely randomly distributed.

From these 1000 reconstructions per cluster, we were able to estimate the ability of our method to detect a cluster of a given mass at a given redshift. This provided us with a measure of the expected completeness with which our method will reconstruct clusters as a function of mass and redshift. We found that at the high end of the mass function $M_{vir}\gtrsim 8\times 10^{14}h^{-1}M_\odot$, we can expect to have a completeness of $\gtrsim 60\%$ up to a redshift of $z_{cl}=0.75$. 

We have shown that our reconstructions show minimal smearing of the structures along the line of sight. The majority of reconstructed clusters lie on only one or two redshift slices ($\Delta z = 0.033 - 0.066$). This represents a significant improvement over linear approaches, where the smearing typically has a \textit{standard deviation} of $\sigma \sim 0.3$. 

These noise realisations were used to estimate the mass and redshift of each cluster by considering the distribution of values across the 1000 noise realisations. We found that the median value provides a robust, unbiased and accurate estimator for the redshift of the cluster. While the error bars associated with the redshift estimates are sometimes fairly large, the redshift estimates themselves are so accurate that they typically give rise to a bias on the mass estimate of less than $5\%$. 

Finally, we showed that with our method, we can directly estimate the mass of the clusters detected.  The most accurate and robust estimate was obtained by taking the median mass contained within the central four pixels of the cluster, but masses obtained at larger radii were consistent with the input at the $1\sigma$ level for most clusters in our sample. 

GLIMPSE therefore allows us to generate density maps from weak lensing measurements of high enough fidelity that we might be able to use these maps to constrain our cosmological model. As an example, \cite{cardone13} have shown that excellent constraints on $f(R)$ modified gravity theories can be obtained by studying the high-mass end of the mass function at high redshift. GLIMPSE now offers the possibility to carry out such a study in weak lensing surveys such as the upcoming Euclid survey. 

Of course, when working with real data, we will have only one realisation of the noise, rather than 1000. However, the redshift and mass estimates shown here, along with their associated errors, can be found using bootstrap resampling methods. Real data also present other challenges: irregularly sampled data, masks, etc. The GLIMPSE algorithm presented here is entirely general, though, and can easily deal with pixellated data whose noise varies pixel-by-pixel. In addition, work is currently underway to modify the implementation of the lensing operator in order to work directly on a shear catalogue, rather than working on binned data, and to apply GLIMPSE to CFHTLens data. 

GLIMPSE represents significant progress in the field of 3D weak lensing density mapping, offering for the first time the ability to reconstruct the cluster field at high accuracy in both redshift and amplitude. Despite being nonlinear, methods exist to constrain the errors in our reconstructions, and these errors are seen, in general, to introduce very little bias on the recovered properties of the clusters detected using this method. This offers the exciting prospect of placing constraints on the high-mass end of the mass function at high redshift using weak lensing surveys.

\section{acknowledgments}
The authors would like to thank Gabriel Rilling, Jerome Bobin, Florent Sureau, Sandrine Pires, Martin Kilbinger and Paniez Paykari for useful discussions. This work is supported by the European Research Council grant SparseAstro (ERC-228261).

\bibliography{refs}

\label{lastpage}

\end{document}